\documentclass[pdflatex,sn-mathphys]{sn-jnl}

\usepackage{amssymb}
\usepackage{amsthm}
\usepackage{framed} 
\usepackage{amsmath,color}
\usepackage{mathrsfs}
\usepackage{graphicx}
\usepackage{epstopdf}
\usepackage{float}
\usepackage{caption}
\usepackage{subcaption}
\usepackage{bm}
\usepackage{bbm}
\usepackage{mathrsfs}
\usepackage{cleveref}

\jyear{2022}%

\theoremstyle{thmstyleone}%
\theoremstyle{thmstyletwo}%

\theoremstyle{thmstylethree}%

\begin{document}

\title[Hygroscopic Phase Field Model ]{Hygroscopic Phase Field Fracture Modelling of Composite Materials}


\author[1]{\fnm{Kit} \sur{Au-Yeung}}
\author[2,3]{\fnm{Adria} \sur{Quintanas-Corominas}}
\author[2]{\fnm{Emilio} \sur{Mart\'{\i}nez-Pa\~neda}}
\author*[1]{\fnm{Wei} \sur{Tan}}\email{wei.tan@qmul.ac.uk}



\affil[1]{\orgdiv{School Of Engineering and Materials Science}, \orgname{Queen Mary University of London}, \orgaddress{\street{Mile End Road}, \city{London}, \postcode{E1 4NS}, \country{UK}}}

\affil[2]{\orgdiv{Department of Civil and Environmental Engineering}, \orgname{Imperial College London}, \orgaddress{\street{Exhibition Rd}, \city{London}, \postcode{SW7 2AZ}, \country{UK}}}

\affil[3]{\orgdiv{AMADE}, \orgname{Universitat de Girona}, \orgaddress{\street{Av. Universitat de Girona 4}, \city{Girona}, \postcode{17003}, \country{Spain}}}




\abstract{This paper investigates the effect of moisture content upon the degradation behaviour of composite materials. A coupled phase field framework considering moisture diffusion, hygroscopic expansion, and fracture behaviour is developed. This multi-physics framework is used to explore the damage evolution of composite materials, spanning the micro-, meso- and macro-scales. The micro-scale unit-cell model shows how the mismatch between the hygroscopic expansion of fibre and matrix leads to interface debonding. From the meso-scale ply-level model, we learn that the distribution of fibres has a minor influence on the material properties, while increasing moisture content facilitates interface debonding. The macro-scale laminate-level model shows that moisture induces a higher degree of damage on the longitudinal ply relative to the transverse ply. This work opens a new avenue to understand and predict environmentally-assisted degradation in composite materials.}




\maketitle

\section{Introduction}\label{sec1}

\begin{figure}[h]
    \centering
    \includegraphics[width=1\textwidth]{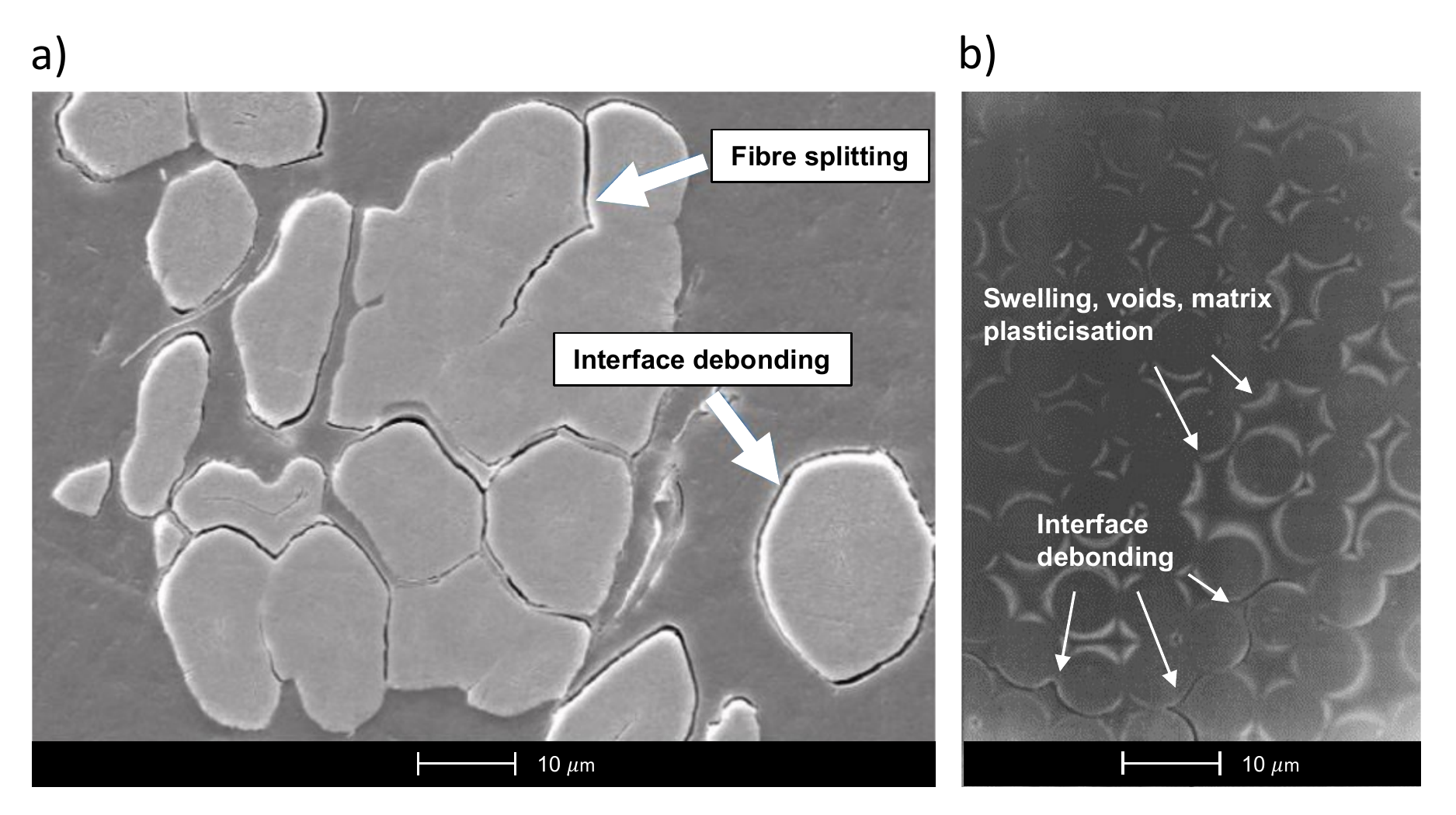}
    \caption{Examples of moisture-induced composite material degradation: (a) moisture-induced debonding and fibre splitting in a flax fibre/epoxy composite \cite{Lu_Maria2018}, and (b) moisture-induced interfacial microcracks in a graphite/epoxy composite \cite{Weitsman2000}.}
    \label{fig:exp_debonding}
\end{figure}

Lightweight composite materials have been widely used in aerospace, wind and marine structural applications due to their high  strength/stiffness to weight ratios and corrosion resistance. However, composite materials are very sensitive to environmental conditions such as moisture and temperature \cite{Barbiere2020}. Moisture contents were found to degrade the mechanical properties of composite materials significantly via fibre-matrix interface debonding \cite{Chilali2018}, hygroscopic expansion \cite{Marwa_Abida2020} and plasticisation of the matrix \cite{Buehler2000, Sateesh2015, Li2020}; see Fig.~\ref{fig:exp_debonding} for examples of moisture-induced composite material degradation. It has been reported that the absorbed moisture in glass fibre epoxy composites results in up to 60\% reduction in tensile strength \cite{Weitsman2000}. Hygroscopic expansion of composite materials is known to cause fibre-matrix interface debonding. This issue becomes more evident in natural fibres such as flax fibre, which are intrinsically hydrophilic and very sensitive to the degree of humidity in the environment \cite{Mohanty2012}. Moisture contents also decrease the glass transition temperature of composite materials, affecting their long-term durability \cite{Yang2004}. Therefore, fundamental understanding of moisture-related degradation mechanisms is essential to enable the development of moisture-resistant composite materials.

A large number of experimental studies have been devoted to the measurement of moisture intake and subsequent material degradation \cite{sathiyamoorthy2020mechanical, chavez2021effect, oguz2021degradation, Buehler2000, Sateesh2015, LeBlanc2016, Lu_Maria2018}. Buehler and Seferis \cite{Buehler2000} found that water absorption led to a decrease in flexural modulus/strength and mode I/II interlaminar fracture toughness of glass fibre reinforced plastics (GFRP). Sateesh et al. \cite{Sateesh2015} reported that the flexural modulus of GFRP decreased by 20-25\% when subjected to a hygrothermal environment for six months. LeBlanc and LaPlante \cite{LeBlanc2016} investigated mixed-mode delamination growth, showing that the delamination toughness decreased by 25-62\% across loading modes. Lu et al. \cite{Lu_Maria2018} reported that the flexural properties of flax-epoxy composites were reduced by up to 20\% when the relative humidity increased to 97\%. However, there are inherent challenges associated with the experimental quantification of micro-scale damage evolution over long periods, and it is often difficult to gain fundamental insight into material degradation mechanisms under coupled mechanical and hygroscopic environmental conditions from laboratory tests. This has triggered an interest in the development of computational models to understand moisture-assisted degradation in composite materials. Some studies are focused on modelling moisture diffusion in composite materials with stationary pre-cracks or cavities. Sinchuk et al. \cite{Sinchuk2018} developed a meso-scale model with Fickian diffusion theory to characterise the moisture diffusion path and local moisture concentration. Gagani and Echtermeyer \cite{Gagani2019} combined a representative volume element (RVE) model with Fickian diffusion to understand transport in composite laminates containing pre-existing cracks and delamination regions. Bourennane et al. \cite{Bourennane2019} used the same methodology and accounted for the role of cavities in the RVE model. Chilali et al. \cite{Chilali2018} simulated the hydro-elastic behaviour of flax fabric-reinforced epoxy composites and obtained the stress distribution at the fibre-matrix interface, although interface debonding was not explicitly modelled. Some authors have also considered the role of growing cracks. For example, Wong et al. \cite{Wong2002} simulated delamination growth in the presence of moisture using moving-node elements. However, this brings in a degree of mesh sensitivity and requires a complex implementation, as re-meshing and an explicit treatment of the boundary conditions at the moving crack faces are needed. Cheng et al. \cite{Cheng2016} developed a hygro-thermal continuum meso-scale model considering damage evolution. Their model solves the diffusion-stress-damage problem in a decoupled fashion and is inherently mesh-dependent. These limitations can be overcome by the use of variational phase field fracture models.\\

The phase field fracture model builds upon Griffith’s thermodynamics principles \cite{Griffith1921} and has been widely used to predict the evolution of cracks in a wide range of materials, including rock-like materials \cite{Fei2021113655,NAVIDTEHRANI2022103555}, ceramics \cite{Carollo20182994,Li2021793}, ductile metals \cite{ambati2015phase,dittmann2018variational}, functionally graded materials \cite{Hirshikesh2019239,Nguyen2021}, porous media \cite{He2022,ZHOU2019169}, shape memory materials \cite{SIMOES2021113504,LOTFOLAHPOUR2023111844}, ice \cite{CLAYTON2022108693,SUN2021101277} and composites \cite{,QUINTEROS2022109788,QUINTANASCOROMINAS2019899}. In addition, it has been successfully employed to model fracture and bridging behaviour in fibre-reinforced composite materials at the microscale level \cite{tan2021phase,tan2022phase}. Moreover, phase field fracture modelling can readily be coupled to equations describing other physical phenomena and thus its use has been particularly popular in addressing multi-physics problems. Examples include hydrogen embrittlement \cite{martinez2018phase,VALVERDEGONZALEZ202232235}, Li-Ion battery degradation \cite{BOYCE2022231119,ZHAO2016428}, thermo-mechanical fracture \cite{MANDAL2021113648,ASURVIJAYAKUMAR2022115096}, and stress corrosion cracking \cite{CUI2022104951,Ansari2021}. However, the use of multi-physics phase field-based formulations to investigate environmentally-assisted degradation in composites materials is an area that remains relatively unexplored. Only very recently two studies have been published in this regard. Arash et al. \cite{Arash2022} presented a phenomenological phase field-based model to investigate the effect of hydrothermal effects on the failure of the behaviour of nanocomposite materials. In this study, the transport problem was not explicitly resolved and, instead, a uniform moisture distribution was assumed. Ye and Zhang \cite{Ye2022} formulated a coupled phase field model that accounted for the interaction of stresses, moisture and crack propagation in composite materials. Their framework incorporated the hygroscopic expansion phenomenon due to moisture diffusion and the viscoelastic behaviour of polymer matrix. They also presented a so-called ``crack filter theory" to regularise the sharp fibre-matrix interface for phase field fracture and moisture diffusion. The crack filter controls the moisture fluxes that can diffuse through the crack. Nevertheless, finding the coefficients for the crack filter functions remains a challenge.\\

In this work, we develop a new phase field-based multi-physics framework for moisture-assisted fracture in composite materials. The framework combines: (i) moisture diffusion, (ii) hygroscopic expansion, (iii) phase field fracture, and (iv) a diffuse interface description of fibre-matrix interface debonding. The potential of the proposed model is demonstrated by addressing three representative benchmark studies spanning the fibre, ply and laminate scales. The results obtained show that the proposed framework can shed new light into the fundamental mechanisms governing environmentally-assisted degradation in composite materials, while also serving as a tool capable of delivering durability predictions over technologically-relevant scales.

\section{Numerical model}
\label{Sec: Numerical model}

\begin{figure}[h]
    \centering
    \includegraphics[width=1\textwidth]{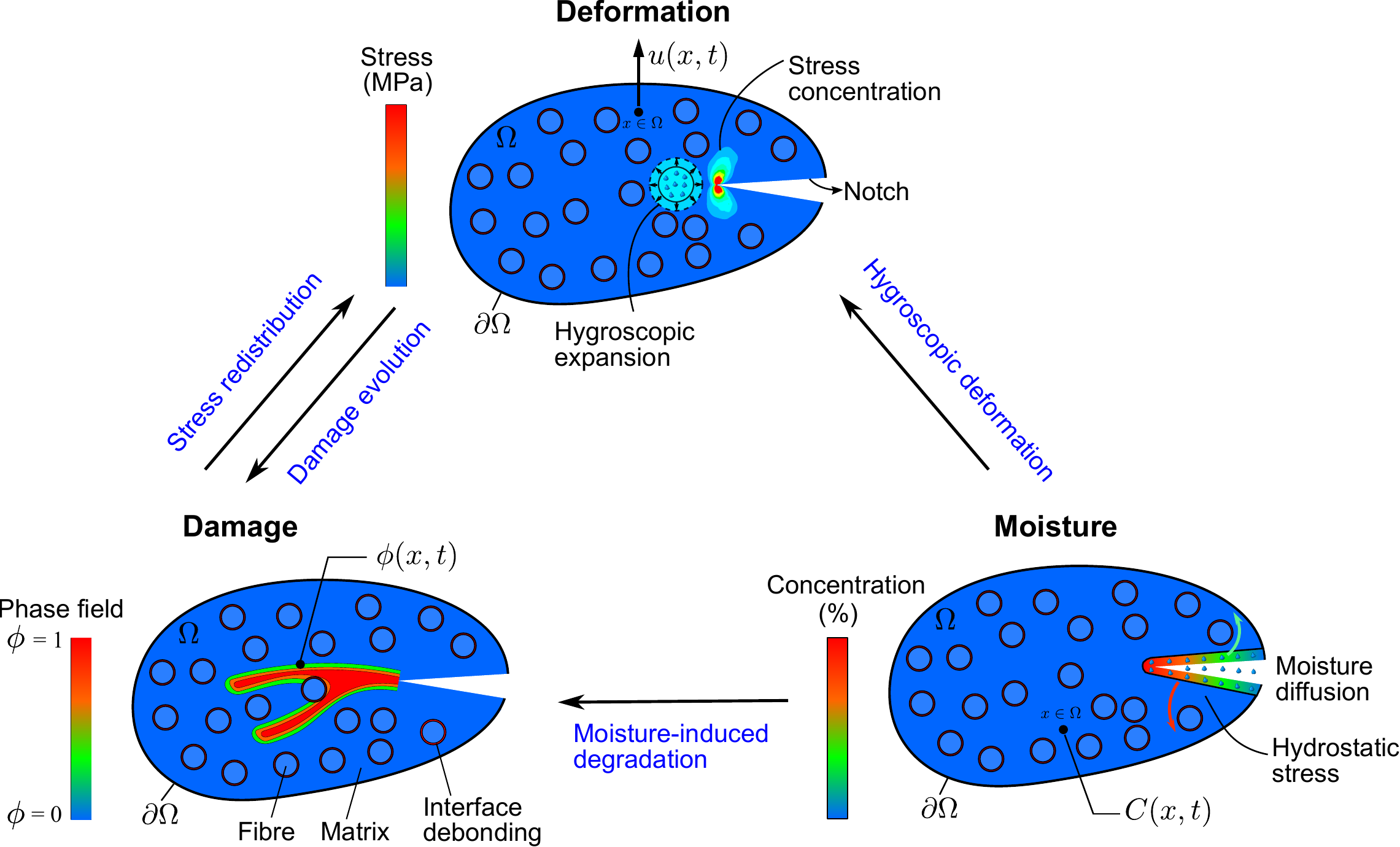}
    \caption{A coupled multi-physics phase field model capturing the interaction between stress, moisture and cracking in the fibre, matrix and fibre-matrix interface.}
    \label{fig:multiphysics_framework}
\end{figure}

In this section, we formulate a new multi-physics phase field model to capture the coupling of moisture diffusion, damage evolution and stress redistribution, as shown in Fig.  \ref{fig:multiphysics_framework}. First, the phase field description of fracture is presented (Section~\ref{Sec:PFfracture}). Then, in Section~\ref{Sec:MoistureDiff}, we proceed to formulate the governing equations for the transport of moisture, including the definition of the hygroscopic strains. A diffused interface theory is subsequently proposed to simulate fibre-matrix debonding (Section~\ref{section:Interface_Indicator}). The coupling of these three main aspects enables a thermodynamically consistent framework that can capture the interplay between mass transport, hygroscopic expansion, crack evolution and stress redistribution. 

\subsection{Phase Field Fracture}
\label{Sec:PFfracture}

We use the phase field model to predict the cracking of fibres, matrix and fibre-matrix interfaces (debonding). The phase field fracture model builds upon Griffith's thermodynamics framework \cite{Griffith1921} - for a crack to propagate, the reduction in potential energy that occurs during crack growth must be balanced with the increase in surface energy resulting from the creation of new free surfaces. In an elastic solid undergoing prescribed displacements, this can be expressed mathematically as follows,
\begin{equation}
\label{eq1}
    \frac{\text{d}\mathscr{E}}{\text{d}A}=\frac{\text{d}\Psi( \bm{\varepsilon}^e)}{\text{d}A}+\frac{\text{d}W_c}{\text{d}A}\
\end{equation}

\noindent where $\mathscr{E}$ is the total energy, d$A$ is an incremental increase in the crack area, $\Psi=\int \psi \, \text{d}V$ is the elastic strain energy, with $\psi$ being the elastic strain energy density, $\bm{\varepsilon}^e$ is the elastic strain tensor, and $W_c$ is the work required to create two new surfaces. The last term in Eq.~(\ref{eq1}) is typically referred to as the critical energy release rate $G_c=\text{d}W_c/\text{d}A$, a material property that characterises the fracture resistance of the solid. For a solid of domain $\Omega$ containing a crack of surface $\Gamma$, Griffith's energy balance can be formulated in a variational form as \cite{Francfort1998}:
\begin{equation}
    \label{eq2}
    \mathscr{E}=\int_{\Omega} {\psi(\bm{\varepsilon}^e)} \, {\text{d}V}+\int_{\Gamma} {G_c} \, {\text{d}\Gamma}
\end{equation}

The evolution of cracks can then be predicted as an exchange between stored and fracture energies. However, minimisation of Eq.~(\ref{eq2}) is computationally challenging due to the unknown nature of $\Gamma$. To overcome this, Griffith's functional can be regularised using the phase field paradigm. The crack-solid material interface is no longer sharp and discontinuous but it is instead smeared over a finite domain and tracked using a phase field order parameter $\phi$. The phase field resembles a damage variable, going from $\phi=0$ in intact material points to $\phi=1$ inside of cracks. Also, following damage mechanics arguments, the stiffness of the solid is degraded by means of a degradation function $g(\phi)={(1-\phi)}^2$. Thus, the regularised functional reads 
\begin{equation}
\label{eq3}
    \mathscr{E}_{\ell}({\bm{\varepsilon}^e,\phi})=\int_{\Omega} (1-\phi)^2 \psi(\bm{\varepsilon}^e) \, {\text{d}V} +\int_{\Gamma} {G_c} \left({\frac{\phi^2}{2\ell}}+{\frac{\ell}{2} \mid{\nabla\phi}\mid^2} \right) \, {\text{d}V}\
\end{equation}
where we have adopted constitutive choices for the degradation function and the crack density function intrinsic to the so-called \texttt{AT2} phase field fracture model \cite{Bourdin2000}. As rigorously proven using Gamma-convergence, the regularised functional $\mathscr{E}_{\ell}$ \eqref{eq3} converges to $\mathscr{E}$ \eqref{eq2} for a choice of phase field length scale $\ell\rightarrow0^+$. The presence of a phase field length scale $\ell$ makes the model non-local and thus ensures mesh objectivity \cite{kristensen2021assessment}. The strong form of the coupled fracture-deformation problem can be readily obtained by taking the variation of $\mathscr{E}_\ell$ with respect to $\bm{\varepsilon}^e$ and $\phi$, noting that the elastic strain tensor is a function of the displacement vector, $\bm{\varepsilon}^e (\mathbf{u})$, and that the stress tensor is given by $\bm{\sigma}=\partial \psi / \partial \bm{\varepsilon}^e$, and applying Gauss' divergence theorem. Accordingly, the strong form equations read
\begin{align}\label{eqn4}
 \nabla \cdot  \left( (1-\phi)^2 \bm{\sigma}_0 \right)  &= \bm{0}   \hspace{3mm} \rm{in}  \hspace{3mm} \Omega \nonumber \\ 
G_{c}  \left( \dfrac{\phi}{\ell}  - \ell \nabla^2 \phi \right) - 2(1-\phi) \, \psi (\bm{\varepsilon}^e) &= 0 \hspace{3mm} \rm{in} \hspace{3mm} \Omega  
\end{align}
where $\boldsymbol{\sigma}_0$ denotes the undamaged stress tensor; $\bm{\sigma}=g(\phi) \bm{\sigma}_0$. 


We adopt a strain energy decomposition to prevent crack growth in compressive stress states. Based on the volumetric-deviatoric split method \cite{Amor2009}, the volumetric and deviatoric strain energies can be subjected to damage but not the compressive volumetric strain energy. Thus, the elastic strain energy is decomposed as ${\psi} = \psi^ +  + \psi^ -$, with
\begin{equation}
\label{eq:VolDevSplit}
    \psi^+ = \frac{1}{2} \lambda \langle \text{tr} \left( \bm{\varepsilon}^e \right) \rangle_+^2 + \mu \left( {\bm{\varepsilon}^e}' : {\bm{\varepsilon}^e}' \right) \hspace{3mm} \text{and} \hspace{3mm}  \psi^- = \frac{1}{2} \lambda \langle \text{tr} \left( \bm{\varepsilon}^e \right) \rangle_-^2
\end{equation}
where ${\bm{\varepsilon}^e}'$ is the deviatoric part of the elastic strain tensor and $\langle x \rangle_{\pm} = 1/2~(x \pm \|x\|)$ are the Macaulay brackets. Also, damage must be irreversible, such that $\dot{\phi} (\mathbf{x}, t) \geq 0$. To enforce this, we introduce a history field $\mathcal{H}$ to store the elastic contribution to the damage driving force. I.e., for a current time $t$, over a total time $t_t$, the history field can be defined as,
\begin{equation}\label{eq:H}
    \mathcal{H} = \text{max}_{t \in [0, t_t]} \, \left( \psi^+ \left( t \right) \right) \, .
\end{equation}
and the phase field evolution equation is reformulated as,
\begin{equation}
G_{c}  \left( \dfrac{\phi}{\ell}  - \ell \nabla^2 \phi \right) - 2(1-\phi) \, \mathcal{H} = 0 \hspace{3mm} \rm{in} \hspace{3mm} \Omega  
\end{equation}

\subsection{Moisture Diffusion}
\label{Sec:MoistureDiff}

Moisture diffusion is governed by Fick's first law. Thus, for a moisture diffusion coefficient $D$, the change in moisture concentration $C$ with time $t$ is given by,
\begin{equation}
\label{eq5}
\frac{\partial C}{\partial t}=D \nabla^2 C
\end{equation}

The driving force for the moisture diffusion is governed by the gradient of the chemical potential $\mu$. The chemical potential of moisture is given by,
\begin{equation}
\label{eq6}
\mu=\mu^0+RT\ln{C}
\end{equation}
where $R$ is the gas constant, $T$ is the absolute temperature and $\mu^0$ denotes the chemical potential on the standard state. Then, the mass flux is defined based on a linear Onsager relationship, such that
\begin{equation}
\label{eq7}
\mathbf{J}=-\frac{DC}{RT}\mathrm{\nabla\mu}=-D\mathrm{\nabla C}
\end{equation}

The mass conservation relates the rate of moisture concentration with moisture flux through the external surface. Using the divergence theorem, the strong form of the balance equation can be obtained, 
\begin{equation}
\label{eq8}
\frac{\text{d}C}{\text{d}t}+\nabla\cdot \mathbf{J}=0
\end{equation}

For an arbitrary scalar field, $\delta C$, the variational form of Eq.~(\ref{eq8}) reads,
\begin{equation}
\label{eq8Variation}
\int_{\mathrm{\Omega}}{\delta C \left( \frac{\text{d}C}{\text{d}t}+\nabla\cdot \mathbf{J} \right) \text{d}V}=0
\end{equation}

Using the divergence theorem and rearranging the equation, the weak form becomes,
\begin{equation}
\label{eq9}
\int_{\mathrm{\Omega}}{\left[\delta C\left(\frac{\text{d}C}{\text{d}t}\right)-\mathbf{J}\cdot\mathrm{\nabla\delta C}\right] \text{d}V+\int_{\partial\mathrm{\Omega}_q}\delta Cq \, \text{d}S}=0
\end{equation}
\newline
where $q=\mathbf{J} \cdot \mathbf{n}$ is the concentration flux leaving the solid through a surface $\partial\mathrm{\Omega}_q$. By inserting the mass flux, Eq.~(\ref{eq7}), into the chemical potential of moisture, Eq.~(\ref{eq9}), the weak form of the moisture diffusion becomes,
\begin{equation}
\label{eq10}
\int_{\mathrm{\Omega}}{\left[\delta C\left(\frac{1}{D}\frac{\text{d}C}{\text{d}t}\right)+\mathrm{\nabla C\nabla\delta C}\right] \, \text{d}V=-\frac{1}{D}\int_{\partial\mathrm{\Omega}_q}\delta Cq \, \text{d}S}
\end{equation}

Changes in time of the moisture concentration lead to the presence of hygroscopic strains $\bm{\varepsilon}^m$. The hygroscopic strain scales linearly with the change of moisture concentration
\begin{equation}
\label{eq12}
\bm{\varepsilon}^m=\alpha \left( C - C_0 \right)
\end{equation}
where $\alpha$ is the moisture expansion coefficient and $C_0$ is the initial moisture concentration in the bulk polymer. Accordingly, the constitutive behaviour of the material incorporates such strains, resulting in the following Cauchy stress definition,
\begin{equation}
\label{eq11}
\bm{\sigma}_0= \mathcal{C} \bm{\varepsilon}^e= \mathcal{C}\left( \bm{\varepsilon} - \bm{\varepsilon}^m \right)
\end{equation}
where $\mathcal{C} $ is the linear elastic stiffness matrix and $\bm{\varepsilon}$ is the total strain tensor. In this regard, it is worth noting that the fibre is assumed to be a transversely isotropic material. 

\subsection{Diffuse interface}
\label{section:Interface_Indicator}

The definition of sharp interfaces complicates the modelling of coupled multi-physics moisture-driven fracture problems. Instead, we choose to use a diffuse interface paradigm to interpolate relevant material properties across the fibre-matrix interface and thus capture fibre-matrix debonding and moisture diffusion across the fibre-matrix interface. To this end, we exploit once again the phase field paradigm to define, as an initial step, a transition zone between the interface and the bulk materials. Thus, a phase field indicator parameter is defined as:
\begin{equation}
\label{eq13}
\left\{\begin{matrix} \mathfrak{d}-\ell_{\mathfrak{d}}^2{\ \mathrm{\nabla}}^2 \mathfrak{d}=0 & \text{in} & \mathrm{\Omega}\\
\mathfrak{d}\left(0\right)=1&\text{on}&\mathrm{\Gamma}\\\mathrm{\nabla \mathfrak{d}}\cdot \mathbf{n}=0&\text{on}&\partial\Omega\\\end{matrix}\right.
\end{equation}

\begin{figure}
    \centering
    \includegraphics[width=1\textwidth]{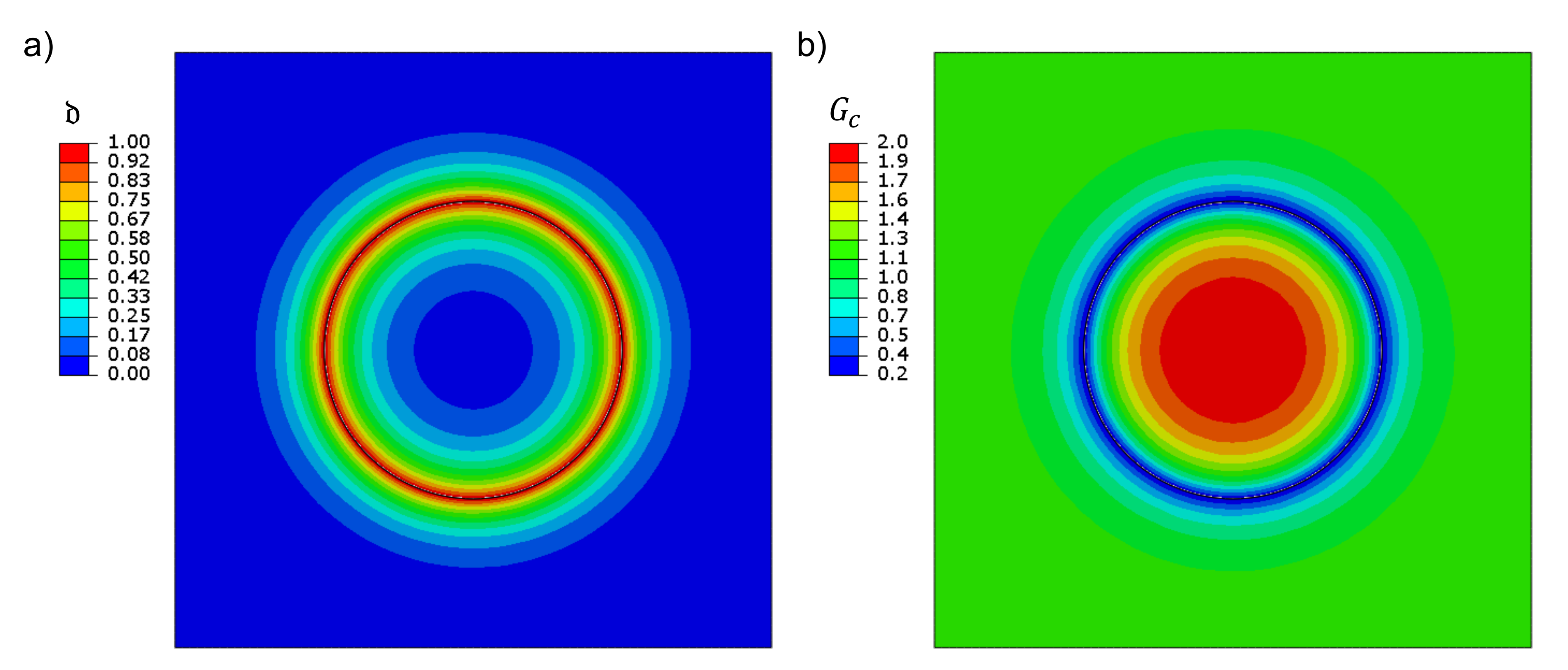}
    \caption{Illustration of the diffused interface approach in a single fibre model: (a) phase field interface indicator $\mathfrak{d}$, and (b) interpolation of the material toughness across the matrix-interface-fibre domains. Subfigure (a) shows how the phase field indicator $\mathfrak{d}$ varies between the interface ($\mathfrak{d}=1$) and the bulk ($\mathfrak{d}=0$) regions, while subfigure (b) illustrates how a material parameter is varied across the fibre (high toughness), interface (low toughness) and matrix regions (intermediate toughness).}
    \label{fig:Gc}
\end{figure}

Similar to the phase field fracture description provided above (see Section~\ref{Sec:PFfracture}), a length scale parameter $\ell_{\mathfrak{d}}$ comes into play to govern the width of the interface region. For simplicity, throughout this work we choose to assume the same magnitude for the phase field fracture length scale $\ell$ and the interface indicator parameter length scale $\ell_{\mathfrak{d}}$. Also, the phase field order parameter $\mathfrak{d}$ is used to interpolate between two limit states: $\mathfrak{d}=0$, corresponding to the bulk, and $\mathfrak{d}=1$, denoting the interface; see Fig.~\ref{fig:Gc}a. An interpolation function $h\left( \mathfrak{d}\right)=\left(1-\mathfrak{d}\right)^{n}$ is used to smoothly transition between the interface, denoted by the superscript $I$, and the bulk, denoted by the superscript $(i)$. The latter is chosen to highlight that the bulk can correspond to two domains: the polymer matrix, $(i)=m$, or the fibre, $(i)=f$. With this in mind, we use this diffuse interface theory to interpolate the material toughness, the diffusion coefficient and the moisture expansion coefficient, 
\begin{align}
\label{eq16}
G_c=h(\mathfrak{d})\left(G_c^{(i)}-G_c^I\right)+G_c^I\\
D=h(\mathfrak{d})\left(D^{(i)}-D_{I}\right)+D_{I}\\
\alpha=h(\mathfrak{d})\left(\alpha^{(i)}-\alpha_{I}\right)+\alpha_{I}
\end{align}

Effectively, this interpolation of material properties prior to the analysis enables simulating fibre-matrix interface debonding, as characterised by the fracture energy of this process, $G_c^I$, and capturing the role that the different nature of the fibre-matrix interface has on moisture transport and hygroscopic swelling. In this regard, the different diffusion coefficients assigned to the interface and the bulk regions capture the different energy barriers inherent to the transport of water molecules through bi-material interfaces. This paradigm is new and provides a simple avenue for incorporating the characteristics of moisture diffusion. In Appendix \ref{app-DiffuseInterface} we provide details of the implementation of this diffuse interface approach into the commercial finite element package \texttt{ABAQUS} by means of user subroutines\footnote{The code developed is made freely available to download at \url{www.imperial.ac.uk/mechanics-materials/codes} and \url{www.empaneda.com/codes}}.

\section{Finite element implementation}
\label{Sec:FEdiscretization}

In the following, we describe the finite element (FE) discretisation of the deformation and fracture problems (Section~\ref{Sec:FEdefPF}), followed by the discretisation of the moisture transport equation (Section~\ref{Sec:FEmoisture}), and the description of the coupled problem and its implementation (Section~\ref{Sec:SolutionScheme}). Common to these sub-sections, the nodal values of the primary fields are interpolated as follows:
\begin{equation}\label{Eq:Discretization}
    \mathbf{u}=\sum_{i=1}^n \bm{N}_i \mathbf{u}_i \,,  \hspace{3mm}
    \phi=\sum_{i=1}^n N_i \phi_i \,,  \hspace{3mm}
    C=\sum_{i=1}^n N_i C_i
\end{equation}
where Voigt notation has been adopted, $n$ denotes the number of nodes and $\bm{N}_i$ are the interpolation matrices - diagonal matrices with the nodal shape functions $N_i$ as components. Similarly, the corresponding gradient quantities are discretised as follows,
\begin{equation}\label{Eq:Discretization2}
    \bm{\varepsilon}=\sum_{i=1}^n \bm{B}^u_i \mathbf{u}_i \, , \hspace{3mm} 
    \nabla \phi=\sum_{i=1}^n \mathbf {B}_i \phi_i\, , \hspace{3mm}  
    \nabla C=\sum_{i=1}^n \mathbf {B}_i C_i\,
\end{equation}
\noindent where $\bm{B}^u_i$ denotes the standard strain-displacement matrices and $\mathbf {B}_i$ are vectors with the spatial derivatives of the shape functions. 

\subsection{FE discretisation of the deformation-phase field problem}
\label{Sec:FEdefPF}

Recall the principle of virtual work and consider the constitutive relations outlined in Section~\ref{Sec: Numerical model}. The weak form for the coupled deformation-phase field fracture problem can be formulated as
\begin{equation}\label{Eq:weak}
  \int_{\Omega} \left[\left( 1 - \phi \right)^2  \bm{\sigma}_0 : \delta \bm{\varepsilon}  -2(1-\phi)\delta \phi \, \mathcal{H}   + G_c \left( \frac{\phi}{\ell} \delta \phi
       + \ell \nabla \phi \cdot \nabla \delta \phi \right) \right]  \, \mathrm{d}V = 0 
\end{equation}

Making use of the FE discretisation given in Eqs.~(\ref{Eq:Discretization}) and (\ref{Eq:Discretization2}), and considering that Eq.~(\ref{Eq:weak}) must hold for arbitrary values of the primal field variables, the residuals can be derived as follows: 
\begin{equation} \label{eq:residualStagU}
    \mathbf{r}_{i}^u =\int_\Omega \left\{ \left[(1-\phi)^{2}+ \kappa \right] {(\bm{B}_{i}^u)}^{T} \bm{\sigma}_0 \right\} \, \mathrm{d}V
\end{equation}
\begin{equation}
    r_{i}^{\phi}= \int_\Omega \left[ -2(1-\phi) N_{i}  \left. \mathcal{H} \right. + G_c \left(\dfrac{\phi}{\ell} N_{i}  + \ell \mathbf{B}_{i}^T \nabla \phi \right) \right] \, \mathrm{d}V 
\end{equation}
with $\kappa$ being a sufficiently small numerical parameter introduced to keep the system of equations well-conditioned when $\phi=1$. We choose to adopt a value of $\kappa=1 \times 10^{-7}$. The components of the stiffness matrices can then be obtained by differentiating the residuals with respect to the incremental nodal variables as follows:
\begin{equation}\label{Eq:Ku}
    \bm{K}_{ij}^{\mathbf{u}} = \frac{\partial \mathbf{r}_{i}^{\mathbf{u}}}{\partial \mathbf{u}_{j}} = \int_\Omega \left\{ \left[(1-\phi)^2+ \kappa \right] {(\bm{B}_i^{\mathbf{u}})}^T \mathcal{C} \bm{B}_j^{\mathbf{u}} \right\} \, \mathrm{d}V  \, ,
\end{equation}
\begin{equation}
    K_{ij}^\phi = \dfrac{\partial r_{i}^{\phi}}{\partial \phi_{j}} = \int_\Omega \left\{ \left[ 2 \mathcal{H} + \frac{G_c}{\ell} \right]  N_{i} N_{j} + G_c \ell \mathbf{B}_i^T \mathbf{B}_j  \right\} \, \mathrm{d}V  \, .
\end{equation}

\subsection{FE discretisation of the mass transport problem}
\label{Sec:FEmoisture}

The residual vector for the moisture transport problem can be readily obtained by discretising Eq.~(\ref{eq10}), given that $\delta C$ is an arbitrary variation of the moisture concentration;
\begin{equation} \label{eq:residualC}
    r_{i}^C =\int_\Omega \left[N_i^T \left( \frac{1}{D} \frac{\mathrm{d}C}{\mathrm{d}t} \right) +  \mathbf{B}_{i}^T \nabla C \right] \, \mathrm{d}V  + \frac{1}{D}\int_{\partial\mathrm{\Omega}_q} N_i^T q \, \mathrm{d}S
\end{equation}
A diffusivity matrix can then be defined,
\begin{equation}
    K_{ij}^C = \int_\Omega \mathbf{B}_i^T \mathbf{B}_j \ \mathrm{d}V  \, .
\end{equation}
Together with a concentration capacity matrix,
\begin{equation} \label{eq:concentrationM}
    M_{ij} = \int_\Omega  {N_i^T\frac{1}{D}{N_j}\mathrm{d}V} 
\end{equation}
and a diffusion flux vector,
\begin{equation} \label{eq:diffusionFlux}
  f_i = -\frac{1}{D}\int_{\partial\mathrm{\Omega}_q} N_i^T q \, \mathrm{d}S
\end{equation}
Subsequently, the discretised moisture transport equation reads,
\begin{equation} \label{eq:moisturesystem}
    \bm{K}^C \mathbf{C} + \bm{M} \dot{\mathbf{C}} = \mathbf{f}
\end{equation}

\subsection{Coupled scheme}
\label{Sec:SolutionScheme}

The deformation, diffusion and phase field fracture problems are weakly coupled. First, mass transport affects the stress field in the fracture process zone via hygroscopic expansion. Secondly, the resulting mechanical fields drive the evolution of the phase field and the reduction in load carrying capacity of the solid. The linearised finite element system,
\begin{equation} \label{eq:coupledScheme}
\left[ {\begin{array}{*{20}{c}}
{{\bm{K}^{\mathbf{u}}}}&0&0\\
0&{{\bm{K}^\phi }}&0\\
0&0&{{\bm{K}^C}}
\end{array}} \right]\left[ {\begin{array}{*{20}{c}}
\mathbf{u}\\
\bm{\phi} \\
\bm{C}
\end{array}} \right]{\rm{ + }}\left[ {\begin{array}{*{20}{c}}
0&0&0\\
0&0&0\\
0&0&\bm{M}
\end{array}} \right]\left[ {\begin{array}{*{20}{c}}
{\dot {\mathbf{u}}}\\
{\dot {\bm{\phi}}}\\
{\dot {\bm{C}}}
\end{array}} \right] = \left[ {\begin{array}{*{20}{c}}
{{\bm{r}^{\mathbf{u}}}}\\
{{\bm{r}^\phi}}\\
{{\bm{r}^C}}
\end{array}} \right]
\end{equation} 
is solved in an incremental manner, using the Newton-Raphson method. The solution scheme follows a so-called \emph{staggered} approach \cite{miehe2010phase}, in that the solutions for the displacement, moisture concentration and phase field problems are obtained sequentially. The implementation is carried out in the finite element package \texttt{ABAQUS} by means of a user element (\texttt{UEL}) subroutine. As described in Appendix \ref{app-DiffuseInterface}, the overall framework is implemented in two steps. In the first step, a \texttt{HETVAL} user subroutine is used to introduce the diffuse interface, by generating a scalar field as the interface indicator. In the second step, the physical simulation is carried out, with a \texttt{UEL} user subroutine being used to define the residuals and stiffness matrices defined above and to assign properties based on the phase field indicator defined in the first stage, as per Eq.~(\ref{eq16}).

\section{Results}
\label{secResults}

We proceed to showcase the abilities of the computational framework developed to predict moisture-induced composite material degradation across the scales. Firstly, in Section~\ref{Sec:ResSingleFibre}, the role of moisture in generating mechanical deformation and damage is investigated using a single fibre model undergoing wetting and drying. Secondly, multiple fibres are considered in a micro-scale model aimed at investigating the role of fibre distribution in moisture diffusion and subsequent material degradation (Section~\ref{Sec:MicroMultiFibre}). Thirdly, in Section~\ref{Sec:MicroCoupled}, we turn our attention to the coupling between mechanical load and moisture transport through a micro-scale single-edge cracked model. Fourthly, a meso-scale model is used to understand the distribution of damage at the ply level and compare model predictions with experiments (Section~\ref{Sec:MesoPly}). Finally, a macro-scale model with 8 plies is studied to investigate the role of hygroscopic swelling on laminate-level structural integrity. Common to all these case studies is the choice of materials adopted. We focus on a composite of epoxy matrix and flax fibre, whose properties are given in Table \ref{tab:material}, together with those of the fibre-matrix interface. Although natural fibres are promising sustainable materials compared to synthetic materials, they are hydrophilic and extremely sensitive to humid environments. 

\begin{table}[h]
    \begin{center}
    \begin{minipage}{275pt}
    \caption{Material properties of the flax fibre, epoxy matrix and epoxy-flax interface.}\label{tab:material}%
    \begin{tabular}{@{}llll@{}}
    \toprule
    Properties & Epoxy & Flax Fibre & Interface\\
    \midrule
    $E_{11}$ (MPa) & 3,600 \cite{Chilali2018} & 31,500 \cite{Chilali2018} & \\
    $E_{22}$ (MPa) & 3,600 \cite{Chilali2018} & 5,100 \cite{Chilali2018} & \\
    $\nu_{12}$ & 0.4 \cite{Chilali2018} & 0.28 \cite{Chilali2018} & \\
    $\nu_{23}$ & 0.4 \cite{Chilali2018} & 0.41 \cite{Chilali2018} & \\
    $G_{c}$ (N/mm) & 1.2 \cite{Ye2022} & 2.1 & 0.213 \cite{Ye2022} \\
    $D$ (mm$^2$/s) & 1.45$\times10^{-6}$ \cite{Joliff2013,Peret2014} & 1.19$\times10^{-6}$ \cite{Celino2013} & 0.8$\times10^{-6}$ \\ 
    $\alpha_{11}$ & 0.6 \cite{Marwa_Abida2020} & 1.06 \cite{Marwa_Abida2020} & 0.1\\ 
    $\alpha_{22}$ & 0.6 \cite{Marwa_Abida2020} & 0.85 \cite{Marwa_Abida2020} & 0.1\\ 
    \botrule
    \end{tabular}
    \end{minipage}
    \end{center}
\end{table}

\subsection{Micro-scale: single fibre model}
\label{Sec:ResSingleFibre}

\begin{figure}[h]
    \centering
    \includegraphics[width=1\textwidth]{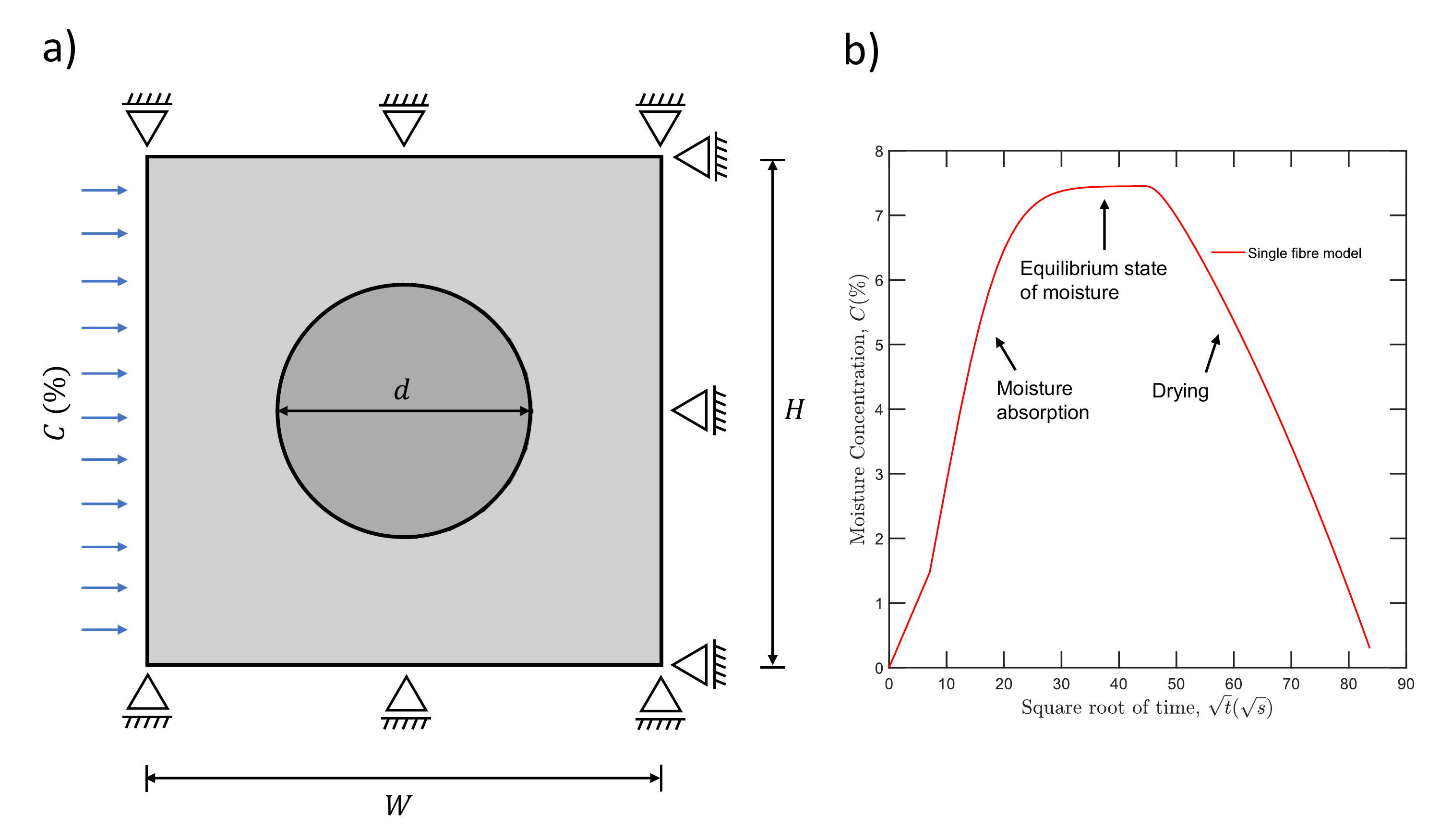}
    \caption{Interfacial damage in a single-fibre model due to hygroscopic swelling: (a) geometry and boundary conditions, and (b) evolution of the moisture content in the centre of the domain as a function of time, showing a moisture absorption stage, followed by an equilibrium plateau and a final drying phase.}
    \label{fig:single_fibre_geometry}
\end{figure}

The single fibre model depicted in Fig.~\ref{fig:single_fibre_geometry}a is first investigated to gain fundamental understanding into interfacial damage due to hygroscopic swelling. The composite dimensions are $H=W=$ 0.02 mm with a fibre diameter of $d=$ 0.01 mm. The vertical displacement is fixed at the top and bottom boundaries, while the horizontal displacement component is fixed at the right boundary. In other words, no mechanical loading is applied and deformation is purely the result of hygroscopic swelling due to moisture gradients. At the left edge, a boundary condition for the moisture concentration is applied. Specifically, a two-stage process is followed, whereby an initial moisture absorption stage is followed by a drying stage. In the first stage the moisture content is prescribed to be equal to 7.45\% \cite{Chilali2018} for 2,000 seconds. This is followed by a second stage where the moisture is reduced to 0\% and the calculation runs for further 5,000 seconds. Initially, the moisture concentration is assumed to be $C(t=0)=0$ in the entire domain. As described in Section~\ref{section:Interface_Indicator}, the interface indicator model is used to capture fibre-matrix debonding, with different work of fractures required for rupturing the epoxy matrix, the flax fibre and the interface (see Table \ref{tab:material}). The entire domain is discretised using 1,862 8-node quadratic elements, with the characteristic element size being equal to 0.0005 mm, two times smaller than the phase field length scale \cite{miehe2010phase}. Fig.~\ref{fig:single_fibre_geometry}b shows the evolution of the moisture content in the centre of the domain (centre of the fibre) as a function of time. It can be seen that initially the moisture concentration increases sharply and eventually a saturation or equilibrium regime is attained, which is followed by a drop in the moisture content after 2,000 seconds.\\

\begin{figure}[h]
    \centering
    \includegraphics[width=1\textwidth]{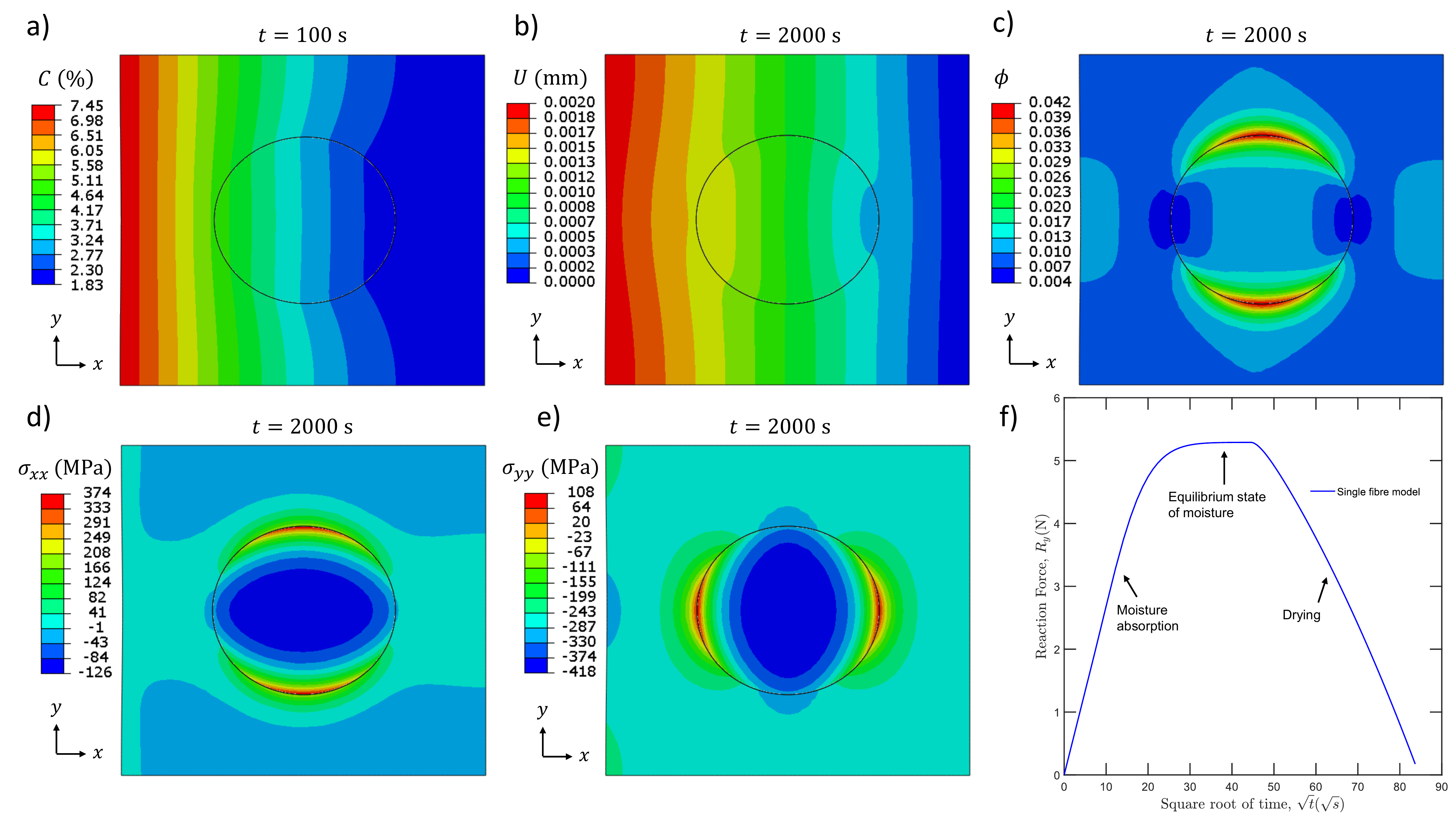}
    \caption{Interfacial damage in a single-fibre model due to hygroscopic swelling: contours of (a) moisture concentration, (b) displacement magnitude, (c) phase field damage, (d) $\sigma_{xx}$ stress component, and (e) $\sigma_{yy}$ stress component. The results are shown for times of $t=100$ s (a) and $t=2,000$ s (b)-(e). Finally, subfigure (f) shows the evolution of the vertical reaction force in time.}
    \label{fig:single_fibre_result}
\end{figure}

The results obtained are shown in Fig.~\ref{fig:single_fibre_result}. Contours of moisture concentration are provided for $t=100$ s in Fig.~\ref{fig:single_fibre_result}a. The $C$ distribution shows a noticeable gradient, which results in hygroscopic strains. The moisture front is not uniform due to the role that the fibre and the matrix-fibre interface play, due to their lower diffusivities. These gradients in concentration eventually disappear as the plateau stage is reached. The impact of moisture diffusion on deformation and damage is given in Figs.~\ref{fig:single_fibre_result}b-f for a time of $t=2,000$ s. Namely, Fig.~\ref{fig:single_fibre_result}b shows the displacement magnitude contours, revealing the impact of hygroscopic expansion. The resulting stresses are given in Figs.~\ref{fig:single_fibre_result}d and e, which provides contours of $\sigma_{xx}$ and $\sigma_{yy}$, respectively. It can be seen that the maximum tensile stresses are attained at the interface due to the material property mismatch. The damage distribution is shown in Fig.~\ref{fig:single_fibre_result}d; damage accumulates at the interface, indicative of fibre-matrix interface debonding. This failure mechanism has been experimentally observed in flax fibre composites \cite{Lu_Maria2018} - see Fig.~\ref{fig:exp_debonding}b. Finally, Fig.~\ref{fig:single_fibre_result}f shows the evolution of the vertical reaction force as a function of time. Three stages can be observed, that resemble the three moisture content stages shown in Fig.~\ref{fig:single_fibre_geometry}b. The load increases as hygroscopic swelling effects increase with moisture diffusion until a more homogeneous $C$ distribution is attained, and the load finally drops during the drying process. Strains and stresses decrease significantly through the drying process but the phase field damage along the interface remains the same due to the irreversibility of damage. 

\subsection{Micro-scale: multi-fibre model}
\label{Sec:MicroMultiFibre}

\begin{figure}
    \centering
    \includegraphics[width=1\textwidth]{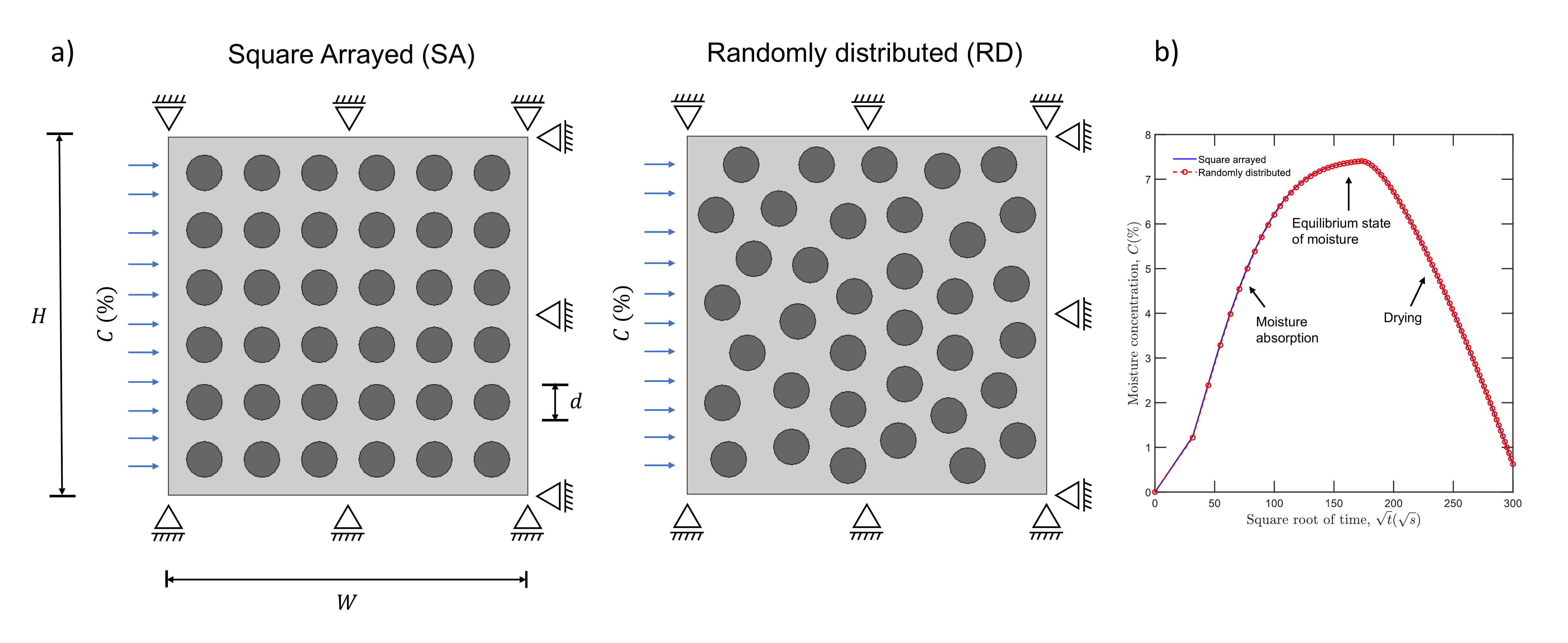}
    \caption{Interfacial damage in a multi-fibre model due to hygroscopic swelling: (a) geometry and boundary conditions for both the Square Arrayed (SA) and Randomly Distributed (RD) cases, and (b) evolution of the moisture content in the centre of the domain as a function of time, showing a moisture absorption stage, followed by an equilibrium plateau and a final drying phase.}
    \label{fig:both_fibre_geometry}
\end{figure}

The modelling framework is subsequently used to shed light into the role of fibre distribution on moisture transport and ensuing material degradation, an area that remains unexplored in the literature. To this end, a micro-scale model of a composite material with multiple reinforcing fibres is adopted. Two scenarios are considered: (i) a Square Arrayed (SA) multi-fibre model, where 36 fibres are uniformly distributed over 6 columns and 6 rows (Fig.~\ref{fig:both_fibre_geometry}a), and (ii) a Randomly Distributed (RD) multi-fibre model, with the same number of fibres as the SA model but randomly distributed (Fig.~\ref{fig:both_fibre_geometry}b). The diffuse interface approach is adopted to introduce the desired properties in all fibre-matrix interfaces and capture multiple fibre-matrix debonding. In both SA and RD scenarios, the composite domain has dimensions of $H=W=0.1$ mm and the fibre diameter equals $d=0.01$ mm. The boundary conditions, sketched in Fig.~\ref{fig:both_fibre_geometry}, involve fixing the vertical displacement in the top and bottom boundaries, and prescribing the horizontal displacement on the right edge. The moisture boundary conditions follow a two-step wet-dry analysis. Thus, for an initially dry sample, a Dirichlet boundary condition $C=7.45$\% is first adopted for a total of 30,000 s, which is sufficient time to reach the equilibrium state. Then, a drying stage begins, where the boundary condition is switched to $C=0$ and the calculation is run for further 60,000 seconds. Fig.~\ref{fig:both_fibre_geometry}b shows the evolution of the moisture content at the centre of the domain, revealing three distinct stages that resemble those observed for the single-fibre analysis (Section~\ref{Sec:ResSingleFibre}). Approximately 60,000 quadratic elements with reduced integration are used to discretise the composite domain, with the characteristic element length being equal to 0.00045 mm, two times smaller than $\ell$ and $\ell_\mathfrak{d}$.\\ 

\begin{figure}
    \centering
    \includegraphics[width=1\textwidth]{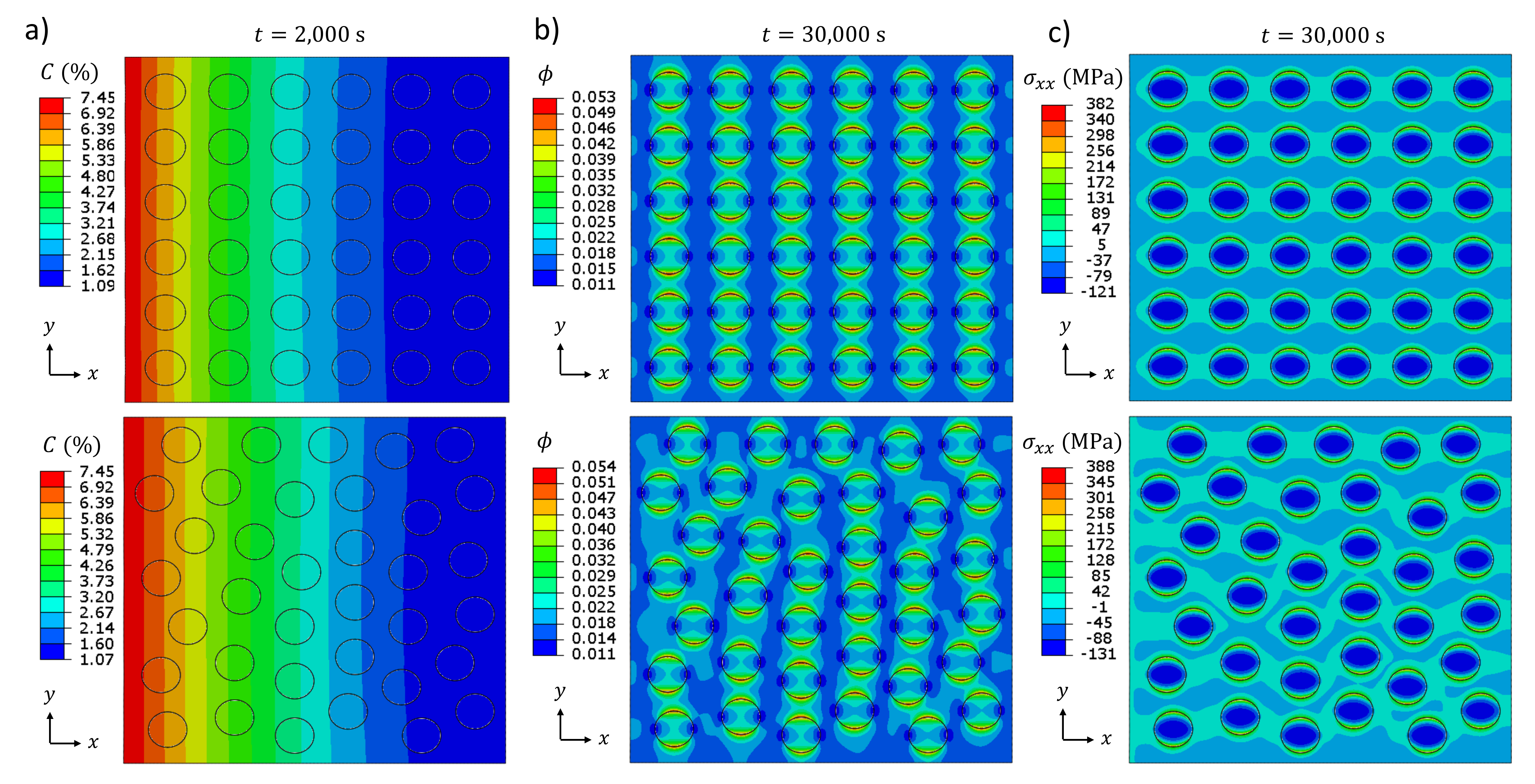}
    \caption{Interfacial damage in a multi-fibre model due to hygroscopic swelling: contours of (a) moisture content after 2,000 s, (b) phase field damage after 30,000 s, and (c) $\sigma_{xx}$ stress component after 30,000 s. The figures in the top correspond to the Square Arrayed (SA) multi-fibre model while those in the bottom have been obtained with the Randomly Distributed (RD) multi-fibre model.}
    \label{fig:both_fibre_result}
\end{figure}

\begin{figure}
    \centering
    \includegraphics[width=0.6\textwidth]{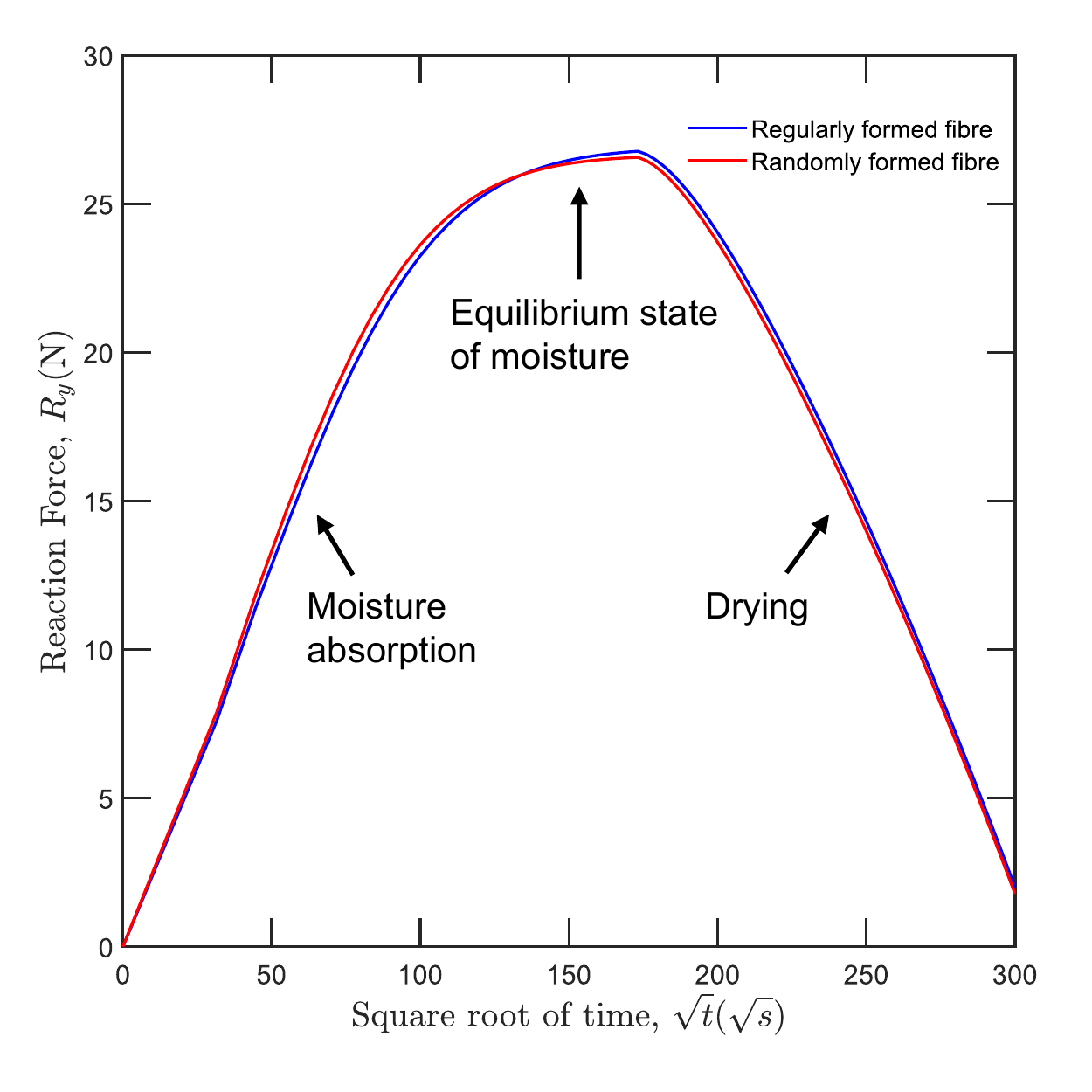}
    \caption{Interfacial damage in a multi-fibre model due to hygroscopic swelling: evolution of the vertical reaction force in time for both the Square Arrayed (SA) and Randomly Distributed (RD) multi-fibre models.}
    \label{fig:RF_result}
\end{figure}

The results obtained are shown in Fig.~\ref{fig:both_fibre_result} in terms of contours of moisture content at $t=2,000$ s (Fig.~\ref{fig:both_fibre_result}a), phase field damage at $t=30,000$ s (Fig.~\ref{fig:both_fibre_result}b), and stress component $\sigma_{xx}$ at $t=30,000$ s (Fig.~\ref{fig:both_fibre_result}c); for both SA and RD fibre models. The results show similar qualitative trends for both scenarios. The trends are also qualitatively similar to those reported for the single-fibre model (Section~\ref{Sec:ResSingleFibre}) albeit some degree of fibre to fibre interaction is observed in terms of stress fields and phase field damage. The force versus time results confirm the impression extracted from the qualitative contours; as shown in Fig~\ref{fig:RF_result} the mechanical responses of the SA and RD fibre models are almost indistinguishable. However, it should be noted that this conclusion is relevant to the material parameters and conditions assumed here. That is, different conclusions might be drawn if the fibre, matrix and interface exhibited higher differences in diffusivity, or if the role of fibre distribution had been explored beyond (e.g., dissimilar volume fractions and fibre radii).\\ 

Following the moisture absorption, a drying stage was applied until $t=$ 60,000 seconds. At the end of the drying stage, the moisture content level decreased from 7.45\% to 0.829\% in the SA model and 0.835\% in the RD model, 0.006\% lesser in the SA model. Fig.~\ref{fig:RF_result}a) shows the result of the vertical reaction force measured from the bottom fixed boundary. During the equilibrium state of moisture, the reaction force of the SA and RD models are 26.04 N and 25.78 N, respectively. The reaction force is in a linear relationship with the hygroscopic expansion, hence the SA model obtained a higher reaction force from a higher moisture content. Fig.~\ref{fig:RF_result}b) shows very similar behaviour in both models during the processes of moisture absorption, equilibrium state, and drying. This is mainly attributed to the relatively uniform distribution of fibre in both models and the close diffusion coefficients of fibre and matrix. It is possible that a non-uniform fibre distribution, for instance, a cluster of fibres, will affect the results significantly. From the above analysis, we can conclude that the uniformly-distributed fibre formation in the matrix will not significantly affect the behaviour of the composite under the hygroscopic expansion effect.

\subsection{Micro-scale: coupled hygroscopic-mechanical model}
\label{Sec:MicroCoupled}

\begin{figure}
    \centering
    \includegraphics[width=0.8\textwidth]{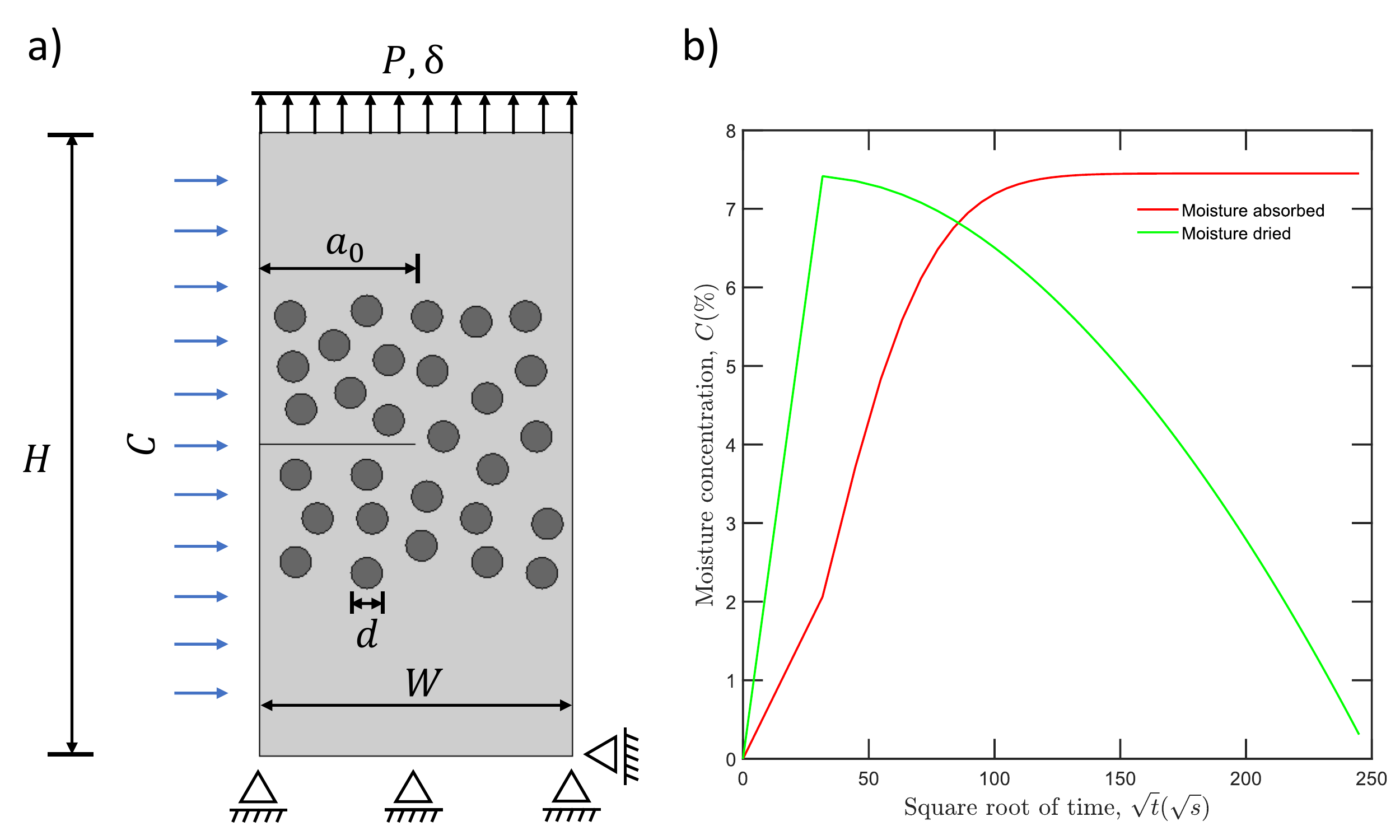}
    \caption{Micro-scale chemo-mechanical analysis of a single-edge cracked plate: (a) geometry and boundary conditions, and (b) evolution of the moisture content in the centre of the domain as a function of time for the cases of moisture absorbed (MA) and moisture dried (MD).}
    \label{fig:flate_geometry}
\end{figure}

A micro-mechanical single-edge cracked plate model is used to explore the influence of moisture diffusion on damage tolerance under the action of both mechanical strains resulting from the application of mechanical load and hygroscopic swelling. The geometry and boundary conditions are given in Fig.~\ref{fig:flate_geometry}a). The composite plate has a width of $W=0.1$ mm, a height of $H=0.2$ mm and a random distribution of fibres with diameter $d=0.01$ mm. An initial crack is introduced geometrically with a length of $a_0={W}/{2}=0.05$ mm. The plate is mechanically loaded by prescribing a vertical displacement at the upper edge, while both vertical and horizontal displacements are prescribed at the bottom boundary. Regarding moisture, three scenarios are considered. In one case (referred to as ``No moisture''), $C=0$ in the entire domain at all times, and damage is driven only by mechanical loading. A second case study, referred to as ``Moisture absorbed'', refers to a sample that is initially dry but that is exposed to a $C=7.45$\% on the left edge for 30,000 s. And a third and final scenario, denoted ``Moisture dried'', assumes that the sample has a uniform, initial moisture content of $C=7.45$\% and then is exposed to a $C=0$\% environment on the left side for 60,000 s (Dirichlet boundary condition). Fig.~\ref{fig:flate_geometry}b shows the moisture content at the centre of the domain for the `Moisture absorbed'' and ``Moisture dried'' case studies. The mechanical-moisture analysis is done in a sequential fashion; that is, the environmental analysis is carried out first, and then a mechanical load is applied where the remote displacement is ramped at a rate of 0.04 mm/s. The entire domain is discretised using 71,390 8-node quadratic elements, with the characteristic element size being equal to 0.00045 mm, two times smaller than the phase field length scale.\\ 

\begin{figure}
    \centering
    \includegraphics[width=1\textwidth]{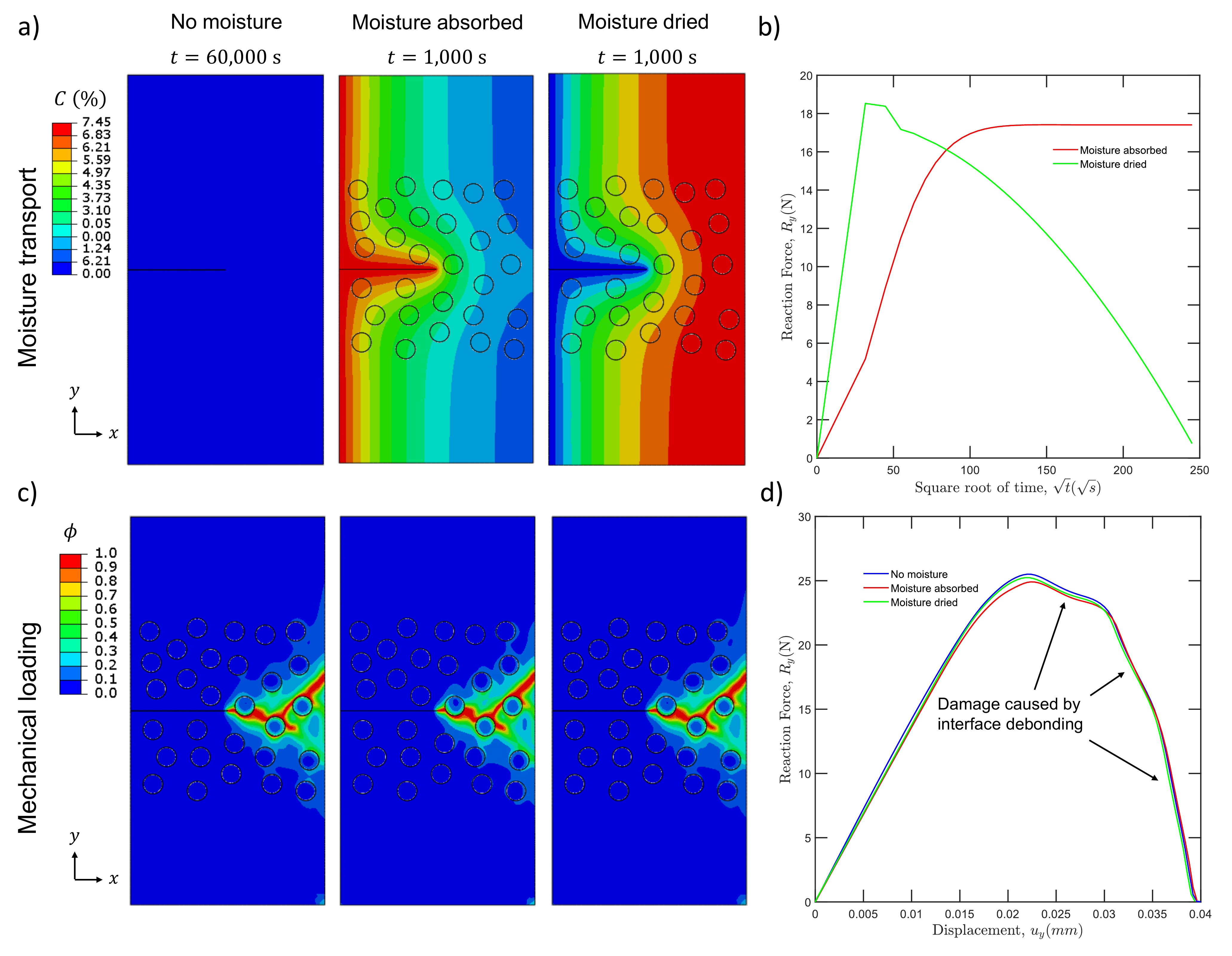}
    \caption{Micro-scale chemo-mechanical analysis of a single-edge cracked plate: (a) moisture concentration contours at relevant times, (b) vertical reaction force versus time before the mechanical load is applied, (c) phase field damage contours showing matrix cracking and interface debonding, and (d) force versus displacement responses. Results are obtained for three case studies: no moisture (NM), moisture absorbed (MA) and moisture dried (MD).}
    \label{fig:flate_result}
\end{figure}

The finite element results obtained are shown in Fig.~\ref{fig:flate_result}. Consider first Fig.~\ref{fig:flate_result}a, showing the distribution of moisture within the domain. In agreement with expectations, the case of moisture absorbed shows a higher content of moisture close to the surface exposed to the environment, while in the moisture dried case there is a reduction in the magnitude of $C$ near the left free surfaces of the plate. Since the crack is introduced geometrically, it has a distinct impact on the moisture contours. Fig.~\ref{fig:flate_result}b shows the reaction force evolution for the moisture absorbed and moisture dried cases. The latter exhibits a drop in the load carrying capacity as a result of the loss of moisture while a monotonic increase is observed for the moisture absorbed case, as represented in Fig.~\ref{fig:flate_result}b, which shows the force versus time responses before mechanical loading is considered. The mechanical and fracture responses obtained after applying a mechanical load are shown in the bottom row of Fig.~\ref{fig:flate_result}. It can be observed that, for both the damage contours (Fig.~\ref{fig:flate_result}c) and the force versus displacement curves (Fig.~\ref{fig:flate_result}d), material behaviour is dominated by mechanical effects and not moisture. This is arguably related to the small degree of damage that is triggered by the moisture, for the material parameters adopted and the environmental conditions considered - see Figs. \ref{fig:single_fibre_result}c and \ref{fig:both_fibre_result}b, the magnitude of $\phi$ remains small throughout the analysis for similar environmental conditions. In all three cases (no moisture, moisture absorbed, moisture dried), matrix cracking and multiple matrix-fibre interface debonding events are observed and these lead to a reduction in load carrying capacity. Fig.~\ref{fig:both_fibre_result}d shows multiple load drop events which are associated with the coalescence of interface debonding and matrix cracks. The peak load of the moisture absorbed model is slightly lower than the one attained for no moisture and moisture dried models but differences are nevertheless small. Considering a moisture concentration-dependent toughness could bring larger differences but this dependency is not clear from the experimental side \cite{Sugiman2016}. 

\subsection{Meso-scale: ply-level model}
\label{Sec:MesoPly}

\begin{figure}[h]
    \centering
    \includegraphics[width=1\textwidth]{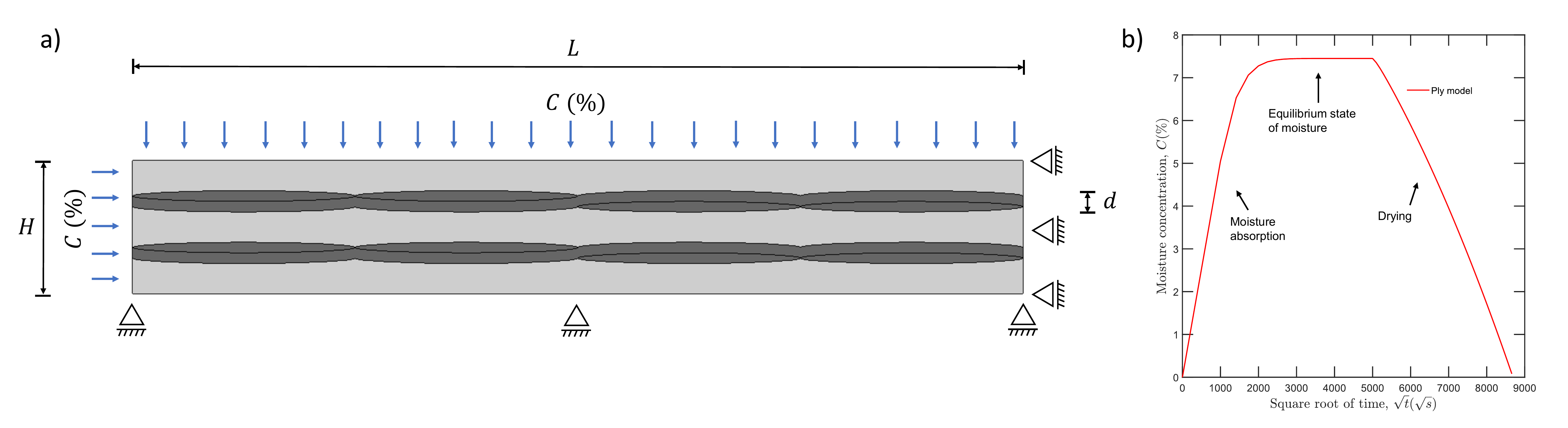}
    \caption{Ply-level model subjected to moisture absorption: (a) geometry and the boundary conditions of the composite, and (b) evolution of the moisture content in the centre of the domain as a function of time, showing a moisture absorption stage, followed by an equilibrium plateau and a final drying phase.}
    \label{fig:ply_geometry}
\end{figure}

A ply-level model is investigated using the same geometry, and boundary conditions as Chilali et al. \cite{Chilali2018} (see Fig.~\ref{fig:ply_geometry}a). The ply dimensions are $L=$ 10 mm and $H=$ 1.5 mm, with the fibre diameter being $d=H\times{V_f}=$ 0.24 mm, where the volume fraction equals $V_f=$ 32\%. The symmetry conditions are located at the bottom and right boundary. A moisture content of 7.45\% is prescribed at the top and left boundaries of an initially dry sample for a total of 25$\times10^{6}$ seconds, which is sufficient to reach the equilibrium state - see Fig.~\ref{fig:ply_geometry}b. After that, the second step reduces the moisture content to 0\% at the same boundary for 50$\times10^{6}$ seconds, as shown in Fig.~\ref{fig:ply_geometry}c. A characteristic element size of 0.013 mm was used, two times smaller than the phase field length scale, which results in a FE model of 107,333 8-node quadratic elements.\\

\begin{figure}
    \centering
    \includegraphics[width=1\textwidth]{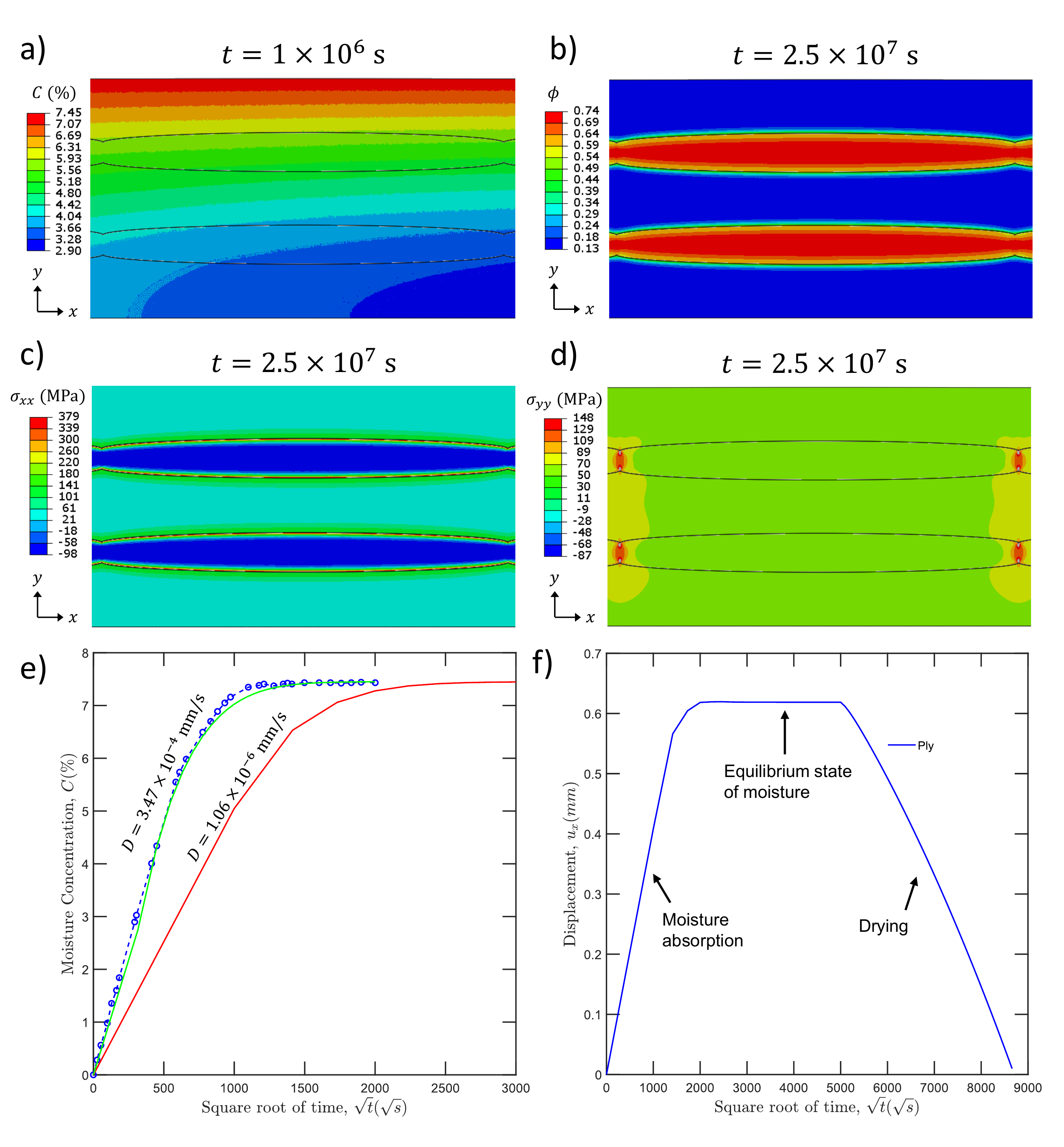}
    \caption{Ply-level model subjected to moisture absorption capture on the middle-left of the domain: (a) moisture concentration contours after $t=1\times10^6$ s, (b) phase field damage contours after equilibrium have been reached ($t=2.5\times10^7$), (c) $\sigma_{xx}$ stress component contours at $t=2.5\times10^7$, (d) $\sigma_{yy}$ stress component contours at $t=2.5\times10^7$, (e) moisture concentration absorption as a function of time, comparison with experiments \cite{Chilali2018}, and (f) evolution of the horizontal expansion of the composite against the square root of time.}
    \label{fig:ply_result}
\end{figure}

The result obtained are shown in Fig.~\ref{fig:ply_result}. The moisture concentration distribution at time $t=1\times10^6$ s (Fig.~\ref{fig:ply_result}a) shows a sharper gradient along the vertical direction, relative to the horizontal one. More importantly, as shown in Fig.~\ref{fig:ply_result}b, the damage resulting from hygroscopic expansion appears to concentrate mainly on the fibre while also spreading slightly into the interface. The internal horizontal stresses resulting from hygroscopic swelling achieve their maximum values at the interface, while notable compressive horizontal stresses are attained inside of the fibre (see Fig.~\ref{fig:ply_result}c). In terms of the $\sigma_{yy}$ stress component, its maximum values are attained at the twist of the fibre, see Fig.~\ref{fig:ply_result}d. In addition, this case study enables us to compare our numerical predictions with experiments. Specifically, as shown in Fig.~\ref{fig:ply_result}e, moisture absorption predictions can be compared with the experimental results by Chilali et al. \cite{Chilali2018}. In turn, the same figure illustrates the sensitivity of moisture transport to the flax fibre diffusion coefficients. The results reveal that, while the value of $D=1.19\times10^{-6}$ mm$^2$/s adopted (see Table \ref{tab:material}) delivers a result that is close to the experimental one, a perfect agreement requires taking a diffusion coefficient of $D=3.47\times10^{-4}$ mm\textsuperscript{2}/s, which is close to the value of 2$\times10^{-4} $ mm\textsuperscript{2}/s experimentally measured by Célino et al. \cite{celino2013characterization} at a relative humidity of 80\%. Finally, Fig.~\ref{fig:ply_result}f shows the horizontal expansion of the laminate due to hygroscopic expansion. During the absorption of moisture, the laminate undergoes a linear horizontal expansion depicted by the movement of the free vertical edge, which ends when the equilibrium state of moisture is reached. At this point, the expansion stops leading to a maximum increase of the laminate length of 0.63 mm. The composite starts to contract during the drying process and the original shape is effectively recovered after 5.0$\times10^{7}$ s. 

\subsection{Macro-scale: laminate-level model}

\begin{figure}[H]
    \centering
    \includegraphics[width=1\textwidth]{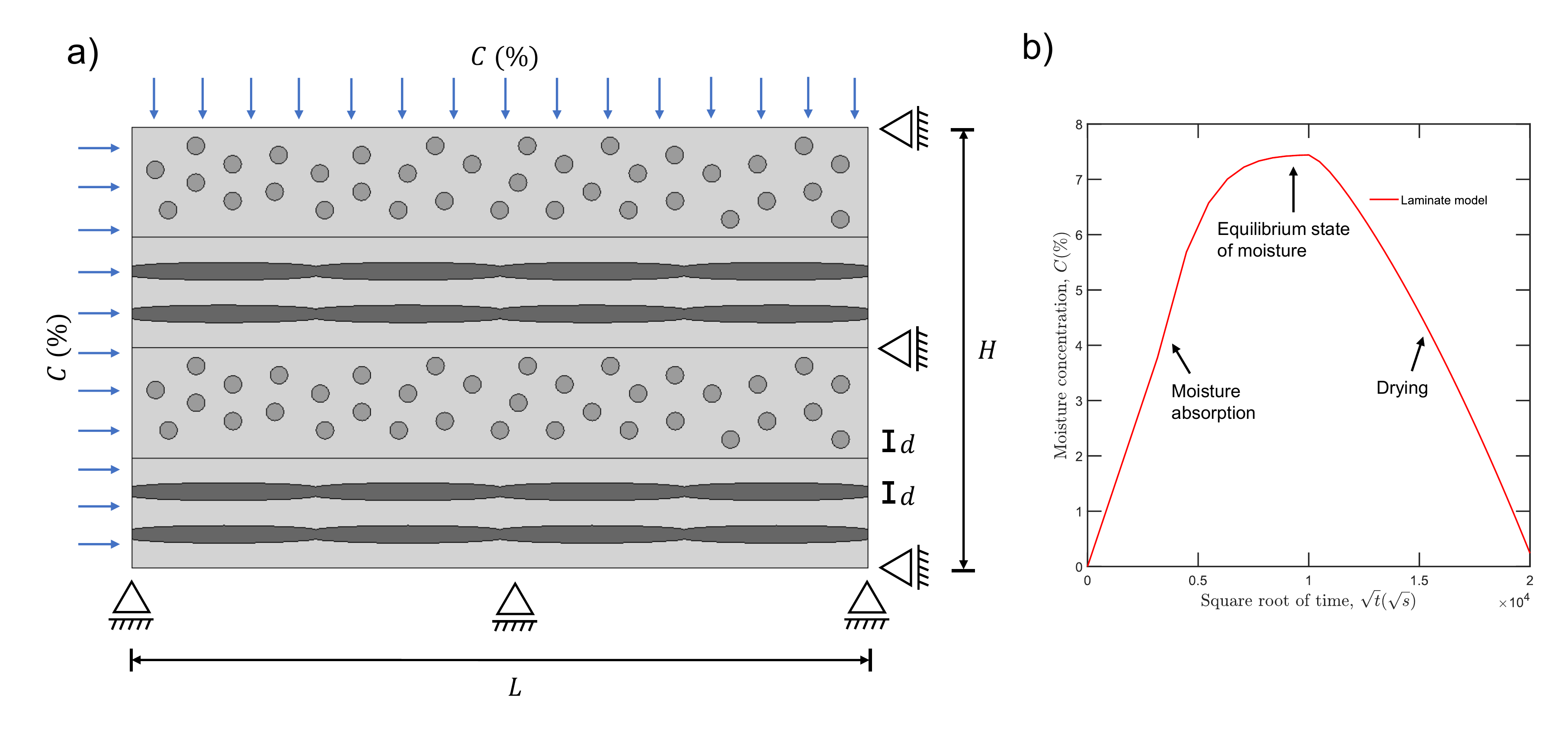}
    \caption{Laminate macro-scale model subjected to moisture absorption: (a) geometry and boundary conditions of the composite, (b) evolution of the moisture content in the centre of the domain as a function of time, showing a moisture absorption stage, followed by an equilibrium plateau and a final drying phase.}
    \label{fig:laminate_geometry}
\end{figure}

The last case study involves a macro-scale, laminate-level model inspired by the work of Chilali et al. \cite{Chilali2018}. Thus, a laminate model comprising 8 plies with a layup of $[0^\circ/90^\circ]{2s}$ and dimensions $L=10$ mm and $H=6$ mm is adopted - see Fig.~\ref{fig:laminate_geometry}a. The fibre diameter is assumed to be $d=0.24$ mm. Due to symmetry conditions at the bottom and right boundaries, only a quarter of the model is simulated. A moisture concentration of 7.45\% is enforced at the top and left boundaries for 1$\times10^{8}$ s which, as shown in Fig.~\ref{fig:laminate_geometry}b, is sufficient to achieve the equilibrium state. This is followed by a drying stage where the moisture concentration is fixed at $C=0$ in the same boundaries (top and left) for further 3$\times10^{8}$s. The entire domain is discretised using 113,565 8-node quadratic elements, with the characteristic element size being equal to 0.013 mm, two times smaller than the phase field length scale. 

\begin{figure}
    \centering
    \includegraphics[width=1\textwidth]{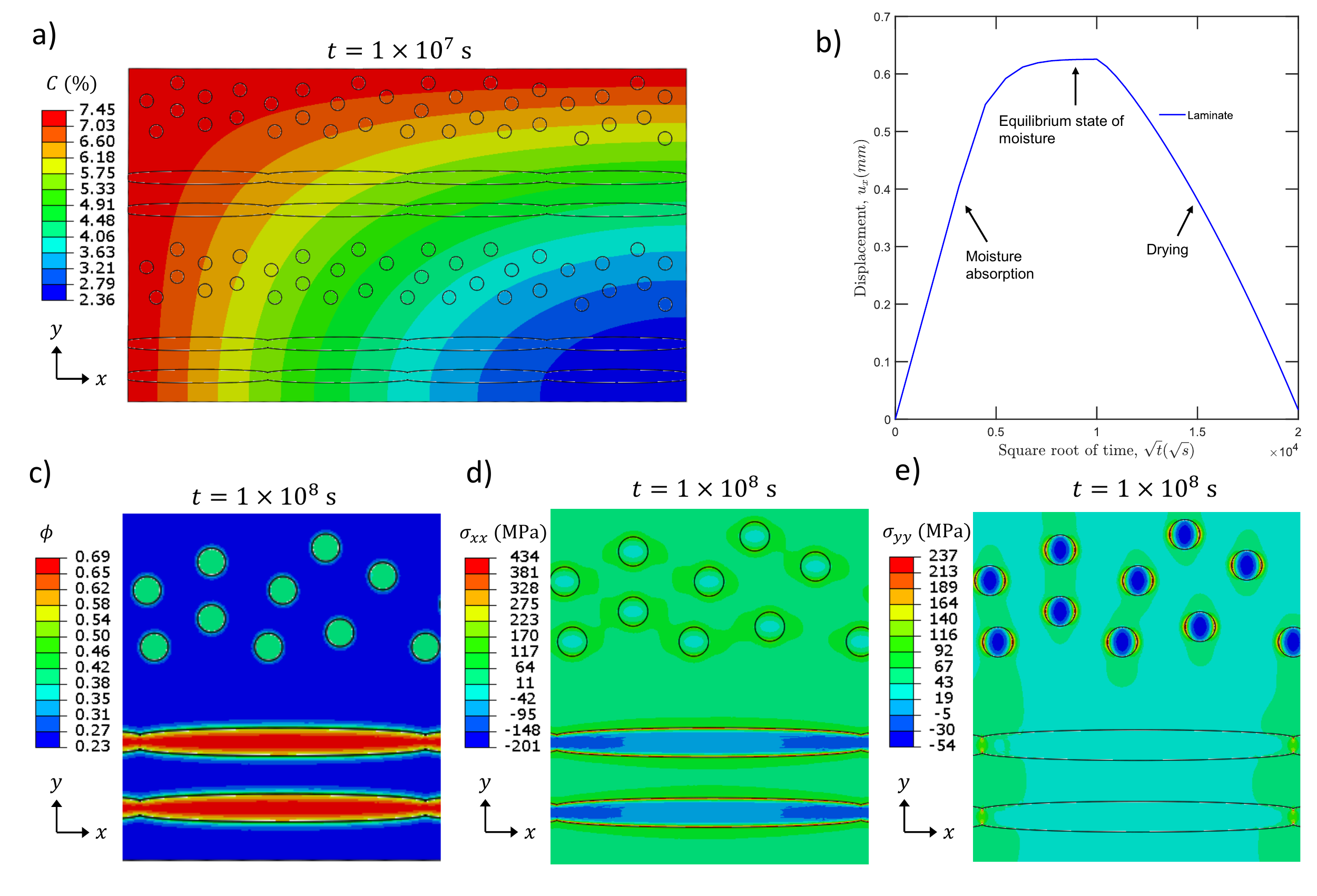}
    \caption{Laminate macro-scale model subjected to moisture absorption: (a) moisture concentration contours after $t=1\times10^7$ s, (b) evolution of the horizontal expansion of the composite against the square root of time, (c) phase field damage contours at $t=1\times10^8$ s, (d) $\sigma_{xx}$ stress component contours at $t=1\times10^8$, and (e) $\sigma_{yy}$ stress component contours at $t=1\times10^8$.}
    \label{fig:laminate_result}
\end{figure}

The computational results obtained are shown in Fig.~\ref{fig:laminate_result} in terms of moisture content contours (Fig.~\ref{fig:laminate_result}a), elongation as a function of time (Fig.~\ref{fig:laminate_result}b), phase field damage contours (Fig.~\ref{fig:laminate_result}c), and $\sigma_{xx}$ (Fig.~\ref{fig:laminate_result}d) $\sigma_{yy}$ (Fig.~\ref{fig:laminate_result}e) stress contours. The moisture contours show gradients in the horizontal and vertical directions, as it could be expected from the boundary conditions adopted. Fig.~\ref{fig:laminate_result}b reveals that the composite elongates by up to 0.61 mm before returning to its original shape as a result of the drying process. Damage appears to concentrate in the longitudinal fibres but a degree of damage is also observed in the transverse ones (see Fig.~\ref{fig:laminate_result}c). The location of the maximum $\sigma_{xx}$ tensile stresses is at the interface between the longitudinal fibres and the matrix (Fig.~\ref{fig:laminate_result}d), while vertical tensile stresses appear to localise at the interface between the matrix and the transverse fibres (see Fig.~\ref{fig:laminate_result}e). The distinctive behaviour observed for the longitudinal and transverse plies is the result of the assumed anisotropy in Young's modulus, Poisson's ratio and hygroscopic expansion coefficient. Overall, the results suggest that moisture will induce a higher degree of damage in the longitudinal plies of composite laminates. 

\section{Conclusions}
\label{sec13}

We have presented a new phase field-based multi-physics framework to model moisture-induced degradation in composite materials. The framework couples the diffusion of moisture, the associated hygroscopic expansion and the mechanical and fracture behaviours of the matrix, fibres and matrix-fibre interfaces. Also, a novel diffuse interface approach is presented to interpolate relevant properties along the fibre-matrix interface. The theoretical framework is numerically implemented using the finite element method and numerical experiments are conducted to gain insight into moisture-material interactions at the micro-, meso- and macro-scales. The main findings are:
\begin{itemize}
    \item The micro-scale analysis of a single-fibre model shows how gradients in concentration lead to hygroscopic strains that result in damage accumulation at the fibre-matrix interface (debonding).
    \item Uniformly distributed and randomised multi-fibre models show that the fibre distribution plays a secondary role for the material parameters and conditions assumed (equal volume fraction and fibre radius).
    \item Micromechanical models involving multi-fibres and combined mechanical and hygroscopic straining show multiple matrix-fibre interface debonding events but a small influence of moisture on the failure process (for the properties and environments considered). 
    \item A meso-scale ply-level model reveals a good agreement with experimental measurements of moisture uptake and predicts that damage localisation will take place mainly along the flax fibre. 
    \item The macro-scale laminate-level model shows that moisture induces a higher degree of damage on the longitudinal ply, relative to the transverse ply, due to the anisotropy of flax fibres.
\end{itemize}

The developed framework constitutes a comprehensive multiscale virtual tool to understand the environmentally-assisted degradation of composite materials. There are multiple avenues of future work. One is to further enhance the efficiency of the calculations by incorporating adaptive mesh refinement techniques, which have shown to be compelling in handling large-scale problems \cite{phansalkar2022spatially,martinez2021adaptive, hirshikesh2021adaptive,pramod2019adaptive}. From a physical viewpoint, other possibilities include accounting for the role of hydrostatic stresses on mass transport and the sensitivity of the mechanical properties to moisture content.

\backmatter

\bmhead{Acknowledgments}
W. Tan acknowledges the financial support from EPSRC New Investigator Award [grant EP/V049259/1] and Royal Society [RGS/R1/231417]. Emilio Mart\'{\i}nez-Pa\~neda acknowledges financial support from UKRI's Future Leaders Fellowship programme [grant MR/V024124/1]. A. Quintanas-Corominas acknowledges financial support from the European Union-NextGenerationEU and the Ministry of Universities and Recovery, Transformation and Resilience Plan of the Spanish Government through a call of the University of Girona (grant REQ2021-A-30).

\begin{appendices}

\section{An Abaqus implementation of a diffuse interface} \label{app-DiffuseInterface}

The generation of the diffuse interface is implemented in \texttt{ABAQUS} making use of a two-step process that involves the definition of an initial phase field indicator $\mathfrak{d}$ distribution and its subsequent incorporation into the physical model using initial conditions and a predefined field variable.\\

The first step involves introducing a phase field distribution so as to interpolate the bulk and interface material properties accordingly. This is an initialisation step that is carried out in the absence of any mechanical or chemical load. To achieve this conveniently in \texttt{ABAQUS}, the analogy between the partial differential equations describing phase field evolution and heat transfer is exploited by means of a \texttt{HETVAL} subroutine \cite{NAVIDTEHRANI2021100050,NAVIDTEHRANI20211996}. Thus, the equation describing the evolution of the indicator parameter $\mathfrak{d}$, Eq.~(\ref{eq13}), can be rearranged as 
\begin{equation}
    \label{eq-DiffuseInterface}
    {\nabla ^2}\mathfrak{d} = \frac{\mathfrak{d}}{\ell_\mathfrak{d}^2}
\end{equation}
where $\ell_\mathfrak{d}$ is the length scale parameter. This equation has the same structure as the heat transfer equation under steady-state conditions when a source term $r$ due to internal heat exists;
\begin{equation}
    \label{eq-HeatTransferSteadyState}
    k\mathrm{\nabla}^2T=r
\end{equation}
Here, $T$ is the temperature field and $k$ is the thermal conductivity. Thus, the $T \equiv \mathfrak{d}$ analogy can be readily attained by defining $k = 1$ and formulating the source term as,
\begin{equation}
   r = \frac{\mathfrak{d}}{\ell^2_\mathfrak{d}} 
\end{equation}

In addition, the \texttt{HETVAL} subroutine requires the definition of the rate of change of $r$ relative to the primary field, which is given by
\begin{equation}
\frac{\partial r}{\partial T} \equiv \frac{\partial r}{\partial \mathfrak{d}} = - \frac{1}{\ell^2_\mathfrak{d}}
\end{equation}

This \texttt{HETVAL} subroutine is used to run a pre-processing steady-state heat transfer analysis in which the Dirichlet boundary condition $T\equiv \mathfrak{d}=1$ is enforced in the set of nodes that describes the (sharp) interface. Then, the physical simulation is conducted, recovering the diffusion interface through the use of a so-called predefined field variable. For this purpose, an initial step with initial conditions is defined, which reads the temperature distribution (output variable \texttt{NT}) from the pre-processing analysis. This field variable is used to interpolate the material properties within a user subroutine. The approach is general, in that it can be combined with any choice of user subroutine (\texttt{UMAT}, \texttt{UELMAT}, \texttt{UEL}) in the second step. A simple example of this implementation, accompanied by documentation, is made available to download at \url{www.imperial.ac.uk/mechanics-materials/codes} and \url{www.empaneda.com/codes}. 
\end{appendices}

\section*{Declarations}

\begin{itemize}

\item \textbf{Conflict of interest}: The authors declare that they have no known competing interests or personal relationships that could have appeared to influence the work reported in this paper.

\item \textbf{Data availability}: All data and codes are available from the author upon reasonable request.

\end{itemize}

\bibliography{sn-bibliography}


\end{document}